%% file: sn-article.tex
\documentclass[pdflatex,sn-apa]{sn-jnl}

\usepackage{graphicx}%
\usepackage{multirow}%
\usepackage{amsmath,amssymb,amsfonts}%
\usepackage{amsthm}%
\usepackage{mathrsfs}%
\usepackage[title]{appendix}%
\usepackage{xcolor}%
\usepackage{textcomp}%
\usepackage{manyfoot}%
\usepackage{booktabs}%
\usepackage{algorithm}%
\usepackage{algorithmicx}%
\usepackage{algpseudocode}%
\usepackage{listings}%
\usepackage{rotating}
\usepackage{pdflscape}
\usepackage{multirow}

\newcommand{\q}[1]{``#1''}
\newcommand{\possessivecite}[1]{\citeauthor{#1}'s (\citeyear{#1})}

\DeclareMathOperator\erf{erf}
\usepackage{tikz,pgfplots} 
\usepgfplotslibrary{groupplots}
\usepackage{pgfplotstable}
\usepgfplotslibrary{external}
\usetikzlibrary{positioning,arrows.meta,shapes}
\pgfplotsset{compat=1.15}

\usepackage{lipsum}
\usepackage{makecell}
\usepackage{longtable}

\theoremstyle{thmstyleone}%
\theoremstyle{thmstyletwo}%

\theoremstyle{thmstylethree}%

\begin{document}

\title[NexGen phase-detection probes for air-water flow experiments]{Developing the next generation of dual-tip phase-detection probes for air-water flow experiments}


\author*[1]{\fnm{Matthias} \sur{Kramer}}\email{m.kramer@unsw.edu.au}

\affil*[1]{Senior Lecturer, UNSW Canberra, School of Engineering and Technology (SET), Canberra,
ACT 2610, Australia,  ORCID 0000-0001-5673-2751}


\abstract{Self-aeration is a fascinating phenomenon that  commonly occurs in high Froude-number flows in natural or human made environments. The most common air-water flow measurement instrument to characterize such flows is the intrusive dual-tip phase-detection needle probe, which identifies phase changes around the needle tips due to a change of resistivity (phase-detection conductivity probe) or light refraction (phase-detection fiber optical probe). Phase-detection conductivity probes are typically custom made for research purposes, with current design dating back to the 1980ies. In the present study, the next generation of dual-tip conductivity probes is developed and validated against a state-of-the-art system. The novel probe design comprises two main features, including (1) a printed circuit board of the sensor's electrodes and (2) a detachable sensor head. These features offer many advantages over the classical needle-type design. For example, the circuit boards can be manufactured with high precision and the sensor head can be easily replaced in case of damage or deterioration. As such, it is anticipated that this relatively cheap and robust design will enable a better repeatability of air-water flow experiments, combined with an enhanced accessibility for the air-water flow research community.}

\keywords{Air-water flow, Dual-tip phase-detection probe, Flow measurement instrumentation, Air concentration}



\maketitle

\section{Introduction}
In hydraulic engineering, high Froude-number free-surface flows at hydraulic structures commonly involve self-aeration, giving rise to complex interactions between air and water phases. The (self-) entrained air is able to alter flow properties of the air-water mixture, leading to flow bulking, drag reduction, cavitation protection, and enhanced gas transfer \citep{Straub1958,Gulliver1990,Falvey1990,Kramer2021Drag}. As such, the current best-practice design approach is based upon air-water flow considerations, comprising a combination of physical and numerical modelling, underpinned by fundamental fluid mechanics judgment \citep{Felder2024}. 

In laboratory- or large-scale physical models of aerated flows, most of the single-phase flow measurement instrumentation becomes limited \citep{JONES197689,Chanson2016}, which is because entrained air bubbles hinder the propagation of light and sound waves. Therefore, specialised air-water flow measurement instrumentation has been developed, of which intrusive dual-tip phase-detection needle probes (PD) constitute the current state-of-the-art \citep{Chanson1995,Felder2016,Tang2022}. These sensors identify phase changes at their tips by a change of electrical conductivity or light refraction \citep{Neal1963,Cartellier1991}, allowing to extract basic air-water flow properties, including air concentrations ($c$), interface frequencies ($F$), interfacial velocities ($u$), and turbulence levels, as well as advanced air water flow properties, such as auto- and cross-correlation integral time scales, and chord sizes \citep{Felder2015}. Because phase-detection optical-fibre probes are relatively expensive and fragile, phase-detection conductivity probes have been the preferred choice for experiments in highly-aerated, high-velocity flows. It is noted that these conductivity probes are mostly custom-made with current design dating back to the 1980ies (Table \ref{table:intro}).

\begin{sidewaystable}
\label{table:intro}
\renewcommand*{\arraystretch}{1}
\include{Figures/tab1}
\end{sidewaystable}

\begin{figure*}[h!]
\centering
\includegraphics[width=\textwidth]{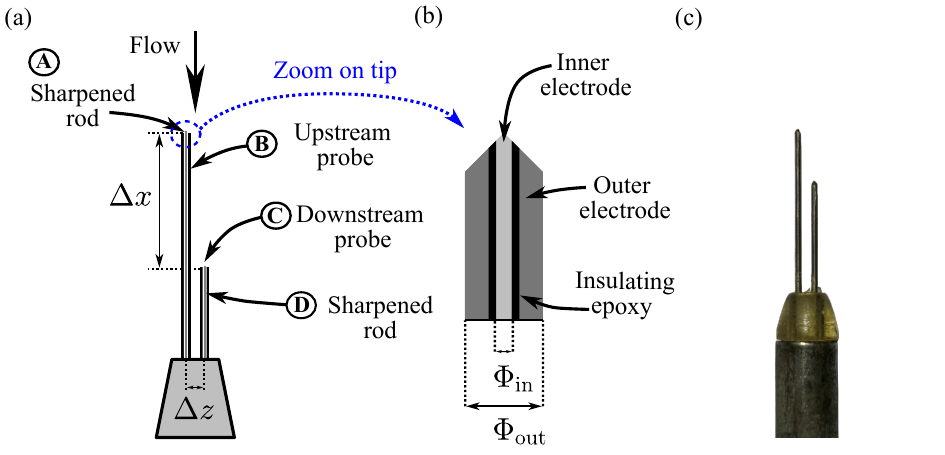}
\caption{State-of-the-art dual-tip needle conductivity phase-detection probe design: (\textit{a}) Schematic of the proposed velocity and air concentration probe by \protect \cite{Cain1981}; the inner electrode diameter $\Phi_\text{in}$ was suggested as  0.02 mm, enabling the detection of small air bubbles; note that the probe-wise ($\Delta x $) and probe-normal ($\Delta z$) tip separation distances were added for convenience (\textit{b}) Zoom on needle-type probe tip with $\Phi_\text{in} = $ inner electrode diameter and $\Phi_\text{out} = $ outer electrode diameter, e.g., \protect 
 \cite{Chanson1995} or \protect 
 \cite{Cummings1996} (\textit{c}) Photograph of the probe head of a state-of-the-art sensor developed at the University of Queensland.}
\label{fig:probe1}
\end{figure*}

In Table \ref{table:intro}, a list of key publications on the development of intrusive conductivity probes for air-water flow experiments is presented. A pioneering effort was made by \cite{Lamb1950}, who were the first to apply an electrical method for measuring air concentrations in flowing waters, replacing previous methods such as mechanical samplers. The conductivity probe of \cite{Lamb1950} used two separated electrodes (separation distance 6 mm) and was advanced by \cite{Keller1972}, who adopted a co-axial probe design, incorporating an inner electrode with diameter $\Phi_\text{in} = 6.35$ mm, and an outer earthed cylindrical electrode. It is noted that the relatively large inner probe tip diameter was selected to allow prototype measurements at the Aviemore Dam (NZ), and that the co-axial alignment of \possessivecite{Keller1972} inner and outer electrodes can be regarded as a direct precursor of today's needle-type electrodes (Fig. \ref{fig:probe1}). In the late 1970ies and 1980ies, the phase-detection probe design was further refined by the research group of Ian Wood at the University of Canterbury in Christchurch \citep{Cain1978,Cain1981,Tan1984,Low1986,Chanson1988}, who were the first to develop a dual-tip needle probe, enabling the measurement of mean interfacial velocities using cross-correlation techniques. In their seminal paper on \q{Instrumentation for Aerated Flow on Spillways}, \cite{Cain1981} proposed a conductivity probe design which has strong resemblance to today's probes (Fig. \ref{fig:probe1}\textit{a}). Subsequently, this dual-tip conductivity needle probe design was sophisticated at the University of Queensland \citep{Chanson1995,Cummings1997}, as summarised in \cite[Table 2]{Chanson2016}, and can be regarded as state-of-the-art design. Later developments by \cite{Thorwarth2008} and \cite{Felder2016} closely follow this approach, while \cite{Tang2022} developed sturdier probes, so-called thick-tip probes, for prototype applications. 

It is noteworthy to mention that smaller electrode diameters were used in the design of dual-tip PD over time (Table \ref{table:intro}, Fig. \ref{fig:probe1}\textit{b}), one reason being the motivation of researchers to reduce the sensor's intrusiveness. While \cite{SHI2023102479} assess intrusive effects to be negligible over a small range of electrode diameters (0.25 mm $ < \Phi_\text{in} <$ 0.64 mm; 0.8 mm $ < \Phi_\text{out} <$ 1.3 mm), it has been shown conclusively that the interaction of dispersed phase particles with the probe tips can lead to significant measurement biases in the estimation of air concentrations and interfacial velocities \citep{VEJRAZKA2010533,Hohermuth2021,PAGLIARA2024104660}. Note that a good overview on the measurement uncertainty of intrusive phase-detection probes is provided in \cite[Table 1]{PAGLIARA2024104660}. Taking the above into consideration, an ideal probe tip should be as small as possible for the sake of measurement accuracy \citep{Felder2024}, which however comes at the expense of fragility, posing a challenge for measurements at prototype scale. In this context, it is emphasized that several production steps of the current approach to manufacturing dual-tip conductivity needle probes involve skilled manual labour, especially when inserting and insulating the inner wire electrode against the outer annular needle, as well as during the sharpening of the tips. As a result, each manufactured probe has individual geometrical features, which may have contributed to aforementioned measurement uncertainties. 

In this study, the next generation of dual-tip conductivity probes is developed, featuring two novel design characteristics: first, the sensor's prongs and electrodes are designed using a printed circuit board (PCB); second, the  sensor head is designed using a detachable connector. It is anticipated that these novel features can overcome some of the aforementioned challenges. For example, the circuit boards can be manufactured with high precision and the fragile sensor head can be replaced in case of damage or deterioration. Hereafter, the development 
of the next generation of dual-tip phase-detection conductivity probes (NexGen PD) is presented ($\S$ \ref{sec:development}), followed by a description of the validation experiment ($\S$  \ref{sec:methods2}) and associated signal processing ($\S$  \ref{sec:methods3}). The NexGen PD is compared against a state-of-the-art PD in $\S$ \ref{sec:results}, demonstrating the suitability of the new sensor for air-water flow experiments. Finally, some limitations are discussed and the importance of the electronic system is emphasized, leading to new research questions formulated in $\S$ \ref{sec:discussion}.

\section{Methods}
\subsection{Development of NexGen phase-detection probes}
\label{sec:development}
\begin{figure*}[h!]
\centering
\includegraphics[width=\textwidth]{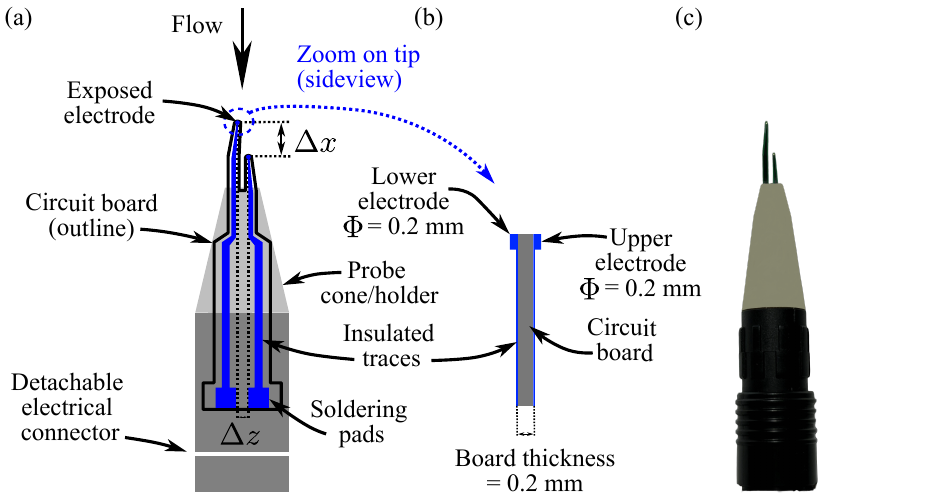}
\caption{NexGen dual-tip phase-detection conductivity probe design: (\textit{a}) Schematic of the probe head, including probe tips and detachable electric connector; $\Delta x =$ probe-wise tip separation distance; $\Delta z =$ probe-normal tip separation distance (\textit{b}) Zoom on the circuit board at the tip; electrode diameters $\Phi$ = 0.2 mm; circuit board thickness = 0.2 mm (\textit{c}) Photograph of the probe head of the NexGen PD.} 
\label{fig:NexGen}
\end{figure*}

The next generation (NexGen) of dual-tip phase-detection probes (Fig. \ref{fig:NexGen}) was developed at the School of Engineering and Technology (SET) at the University of New South Wales (UNSW). The developed dual-tip phase-detection probe system consists of three components, including (1) a probe head with leading and trailing tips, (2) a probe support arm, and (3) an electronic box. The probe head serves as holder for the circuit board, which forms a leading and a trailing tip, separated by $\Delta x = 5$ mm and $\Delta z = $ 1 mm, respectively (Fig. \ref{fig:NexGen}\textit{a}).

Each tip contains two exposed circular electrodes ($\Phi = 0.2$ mm), located on the upper and lower side of the circuit board, which has a thickness of 0.2 mm  (Fig. \ref{fig:NexGen}\textit{b}). Insulated traces lead from the exposed electrodes to the electrical connector, which were soldered via larger base pads to the connector's sockets. It is emphasized that the two exposed electrodes of each tip have identical dimensions, which is different to the state-of-the-art needle-type PD, where the exposed part of the outer electrode is typically much larger than the exposed part of the inner electrode, compare Fig. \ref{fig:probe1}\textit{b}. Note that the outer electrode of needle-type probes is sometimes insulated, as done in \cite{Tang2022}, but to the best knowledge of the author, the majority of outer electrodes of current needle-type probes do not carry a second insulation layer. As such, the NexGen PD should hypothetically be able to detect a larger amount of dispersed phase particles, as both electrodes contribute to their detection, as describe in the following: if both electrodes are surrounded at the same time by the same fluid phase (here: air or water), the voltage output is proportional to the conductivity of the respective fluid phase, which can be translated into an instantaneous air concentration. In flow regions with a distinct carrier phase, i.e., water in the bubbly flow region or air in the droplet region, the output of the NexGen PD sensor corresponds to the conductivity of the carrier phase, unless one of the two electrodes pierces the respective dispersed phase. 

In relation to the probe support arm and the electronic box, the former had a standard L-shape with horizontal and vertical dimensions of 150 mm and 460 mm, respectively. The receptacle of the electrical connector was glued into the end section of the horizontal part of the support arm, while an insulated copper cable (3 m length) was waterproofed with heatshrink at the end of the arm's vertical part. For the measurements in the present study, an electronic box/air bubble detector was newly designed. This new electronic circuit was based on the measurement of current (instead of voltage), which is less prone to external noise, further providing raw analogue output signals of the probe tips. Some more considerations on the different electronic circuits are discussed in $\S$ \ref{sec:hysteresis}, while the complete circuit layout of the present development is not presented here, due to intellectual property protection. 

\subsection{Validation experiment and flow conditions}
\label{sec:methods2}
\begin{figure*}[h!]
\centering
\includegraphics[width=\textwidth]{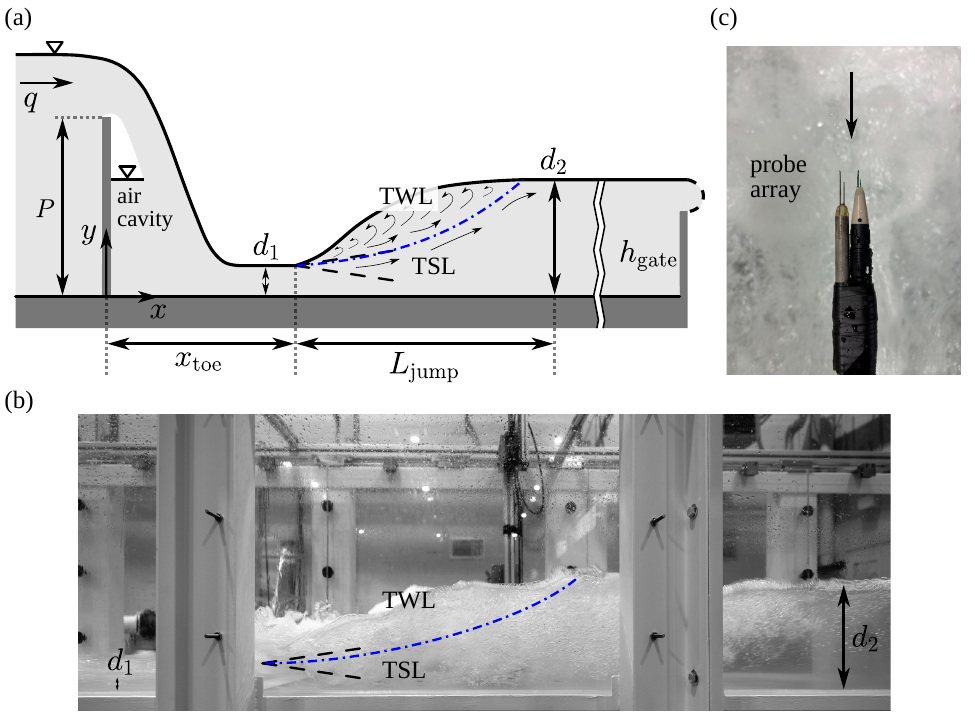}
\caption{Experimental set-up: (\textit{a}) Hydraulic jump downstream of a sharp crested weir at UNSW Canberra; TSL = Turbulent Shear Layer; TWL = Turbulent Wavy Layer, corresponds to recirculation zone/jump roller; (\textit{b}) Snapshot of the hydraulic jump during experimental campaign (image courtesy of Elena Pummer);(\textit{c}) Phase-detection probe array used for comparative air-water flow measurements.} 
\label{fig:experimentalsetup}
\end{figure*}

To test and validate the NexGen PD, new air-water flow experiments were conducted at UNSW Canberra within a hydraulic jump, downstream of a linear sharp crested weir (Figs. \ref{fig:experimentalsetup}\textit{a,b}). The weir height and the discharge were set to $P=0.4$~m and $q = 0.11$~m$^2$/s, the latter controlled by a centrifugal pump with variable speed drive. Note that the width of the weir corresponded to the channel width of 0.6 m. Downstream control of the 9 m long channel was achieved using a sluice gate with a height of $h_\text{gate} = 0.091$ m (Fig. \ref{fig:experimentalsetup}\textit{a}), corresponding to $h_\text{gate}/P = 0.23$. The conjugate flow depths, inflow Froude number, and jump length were $d_1 = 0.045$ m,  $d_2 = 0.216$ m, $Fr_1 = 3.61$, and $L_\text{jump} = 1.82$ m, respectively. All experimental flow conditions are summarized in Table~\ref{table:conditions}, additionally including the dimensionless location of the jump toe position ($x_\text{toe}/P$) as well as conjugate depth ratio $(d_2/d_1)$, which had a deviation of less than 4\% from the well-known Belanger equation \citep{Belanger1849}. 

For the comparative air-water flow measurements, a phase-detection probe array, consisting of a NexGen PD and a state-of-the-art PD (Fig. \ref{fig:experimentalsetup}\textit{c}), was used to simultaneously measure five vertical profiles at dimensionless streamwise locations $(x - x_\text{toe})/d_1 =$ 3.1, 5.9, 8.3, 10.9, and 13.4 along the jump's centreline. The NexGen PD system corresponded to the developments described in the previous $\S$ \ref{sec:development}. The state-of-the art PD system was a dual-tip needle conductivity probe with an inner electrode made of platinum wire ($\Phi_\text{in} = 0.25$ mm) and an outer stainless steel needle ($\Phi_\text{out} = 1.0$ mm), having tip separation distances of $\Delta x = 6.5$ mm and $\Delta z = 2.2$ mm, respectively. The complete system was borrowed from the University of Sydney for this research, while it was originally manufactured at the University of Queensland. The probe was excited by an electronic box/air bubble detector A25240, of which a circuit diagram and more detailed descriptions are provided in \cite[Appendix E]{Chanson1988} and in \cite[Appendix G]{Cummings1996}, respectively. It is noted that the air bubble detector A25240, or modified versions thereof, are currently used by several groups within the air-water flow research community \citep{Chanson1995,Felder2016}.

The vertical position of the probe array was controlled with a linear stepper motor and the probes were sampled at a sampling rate of 20 kHz for 60 s using a CompactDAQ system from National Instruments. Throughout the experimental measurement campaign, it was carefully ensured that boundary conditions, such as flow rate, flow depths, and jump toe position, remained identical.

\input{Figures/tab2.tex}

\subsection{Signal processing}
\label{sec:methods3}
Phase-detection signal processing was performed following the best-practice guidelines outlined in \cite{Kramer2020}. For the binarization of the voltage output of the probe tips into a time series of instantaneous air concentrations, a single threshold technique based on the intermodal range of the tip's  signal probability density functions was used \citep{Cartellier1991,Felder2015,Kramer2020}
\begin{equation}
\label{eq:singlethreshold}
\frac{S_1(t) - m_a}{(m_w - m_a )} \, \begin{cases}
 \geq \mathcal{T} \rightarrow c_1(t) = 0, \\
< \mathcal{T} \rightarrow c_1(t) = 1,
\end{cases}
\end{equation}
where $S_1$ is the voltage signal of the leading tip (subscript 1), $t$ is the time, $m_a$ and $m_w$ are the air and water modes of the signal's probability density function (in voltage), respectively, and $\mathcal{T}$ is the dimensionless single threshold, bounded by zero and one,  whose selection is discussed in the next section. In a steady stationary flow, 
the time averaged air concentration $\overline{c}$ corresponded to the mean value of the instantaneous air concentration time series, defined as 
\begin{equation}
\overline{c} = \frac{1}{T} \int_{t = 0}^T c_1 \, \text{d} t,
\label{eq:timeaverage}
\end{equation}
with $T = 60$ s being the sampling duration. For comparative reasons, the interfacial frequency $F$ was defined as half the number of air-water interfaces pierced by leading tip over the sampling duration. 
Velocity estimations were performed using the Adaptive Window Cross-Correlation technique \citep{Kramer19AWCC,Kramer2020,Kramer20Practicescomment}. The AWCC technique segments the instantaneous air concentration signals into small windows, considering a reduced number of particles for cross-correlation analysis. The quality of estimated pseudo-instantaneous velocities was monitored using the cross-correlation coefficient and the secondary peak ratio \citep{Kramer2020}, implying that velocity data was discarded when insufficient correlation was obtained. In agreement with recommendations of \cite{Kramer2020}, the number of particles/interface pairs for AWWC analysis was set to $n_p =6$.

\section{Results}
\label{sec:results}

\subsection{Probe signals and their probability distributions}

\begin{figure*}[h!]
\input{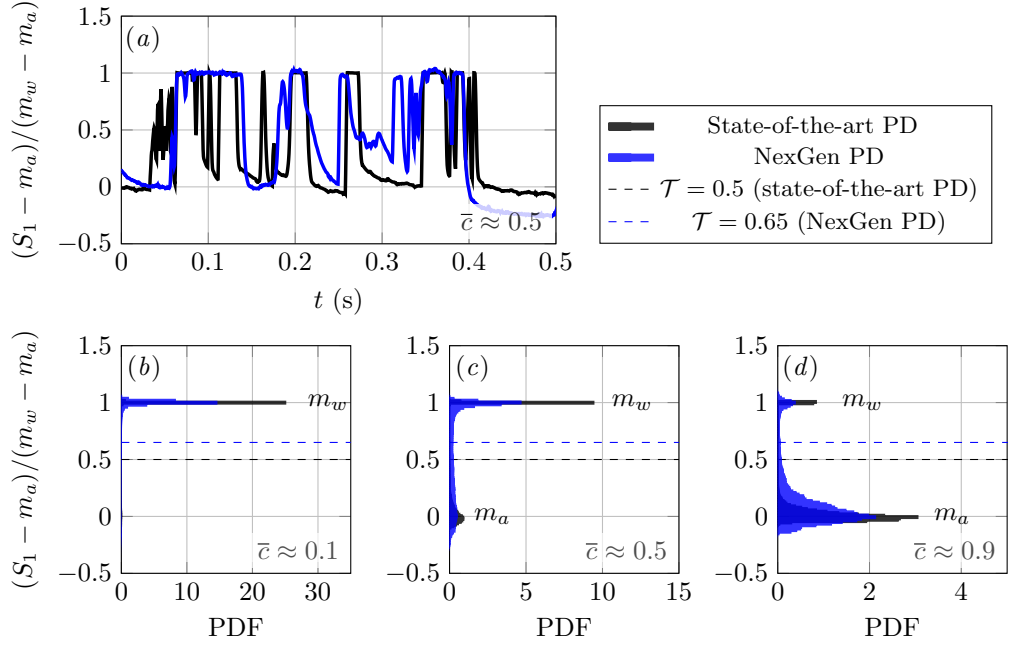}

\vspace{-0.5cm}

\centering
\input{Figures/fig4.tex}
\caption{Leading tip signals ($S_1$) of a state-of-the art PD and the NexGen PD at a streamwise location $(x-x_\text{toe})/d_1$ = 5.9: (\textit{a}) Representative time series of PD signals at $y/d_1 = 2.94$ with $\overline{c} \approx 0.5$; note that the ordinate is normalised using the modes $m_a$ and $m_w$; (\textit{b}) Probability density function estimates at $y/d_1 = 0.94$ with $\overline{c} \approx 0.1$;  (\textit{c}) at $y/d_1 = 2.94$ with $\overline{c} \approx 0.5$; (\textit{d}) and at  $y/d_1 = 3.72$ with $\overline{c} \approx 0.9$.}
\label{fig:raw}
\end{figure*}

Representative signals of the leading tips of the two phase-detection probes at a streamwise location $(x-x_\text{toe})/d_1 = 5.9$ are shown in Fig. \ref{fig:raw}, including a short fragment of a recorded time series (Fig. \ref{fig:raw}\textit{a}) as well as three probability density functions (PDF, Figs. \ref{fig:raw}\textit{b} to \textit{d}). For this representation, three different air concentrations $\overline{c} = 0.1$, 0.5, and 0.9 were selected, which is because the mixture flow depths $y_{50} = y(\overline{c} = 0.5)$ and $y_{90} = y(\overline{c} = 0.9)$  are important physical air-water flow parameters, oftentimes used for the characterization of the air concentration distribution \citep{Kramer2023,Kramer2024}.  
The corresponding dimensionless elevations were $y/d_1 = 0.94, 2.94$, and 3.72, respectively, located in the turbulent shear layer and the roller region of the hydraulic jump. 

Figure \ref{fig:raw} demonstrates an overall good agreement between the state-of-the-art PD and the NexGen PD  for the  signals and their distributions, while it is noted that the probability density functions of the NexGen are wider, implying that potentially more interfacial information is recorded between the two modes of the PDFs. The reasons for this observation are believed to be manifold. First, the electronic box/air bubble detector A25240 of the state-of-the-art PD contains a Schmitt-Trigger, which converts the raw analogue signal into a digital signal using an upper and a lower threshold. While these two thresholds have not commonly been reported in air water flow literature, an exemption being the thesis of \cite{Cummings1996}, it is evident that the Schmitt-Trigger removes some physical information from the raw signal. In this context, it is noted that the electronic circuit of the NexGen PD delivers analogue raw signals without prior filtering, which is believed to be advantageous. Second, as outlined in $\S$ \ref{sec:development}, both electrodes of the NexGen PD contribute to the detection of interfaces, and as such, the NexGen PD is able to capture more interfacial information. Following from the above, observed differences between the signals of both systems are expected, which raises the question on the selection of the binarization threshold $\mathcal{T}$ [Eq. (\ref{eq:singlethreshold})]. 

In line with previous studies, amongst others \cite{Felder2015}, \cite[Table 2]{Chanson2016}, and  \cite{SHI2023102479}, the single threshold for signal binarization of the state-of-the-art PD was set to $\mathcal{T} = 0.5$, which was also deemed important for comparability with previous literature. For the NexGen PD, the single threshold  was selected as $\mathcal{T} = 0.65$, which was because the PDFs had a local minimum around this value (c.f. Figs. \ref{fig:raw}\textit{c,d}), and because a good agreement of time-averaged air concentrations between the NexGen PD and the state-of-the-art PD was achieved using this setting.

\newpage
\subsection{Air concentration, interface frequency, and chord times}

\begin{figure*}[h!]
\centering
\input{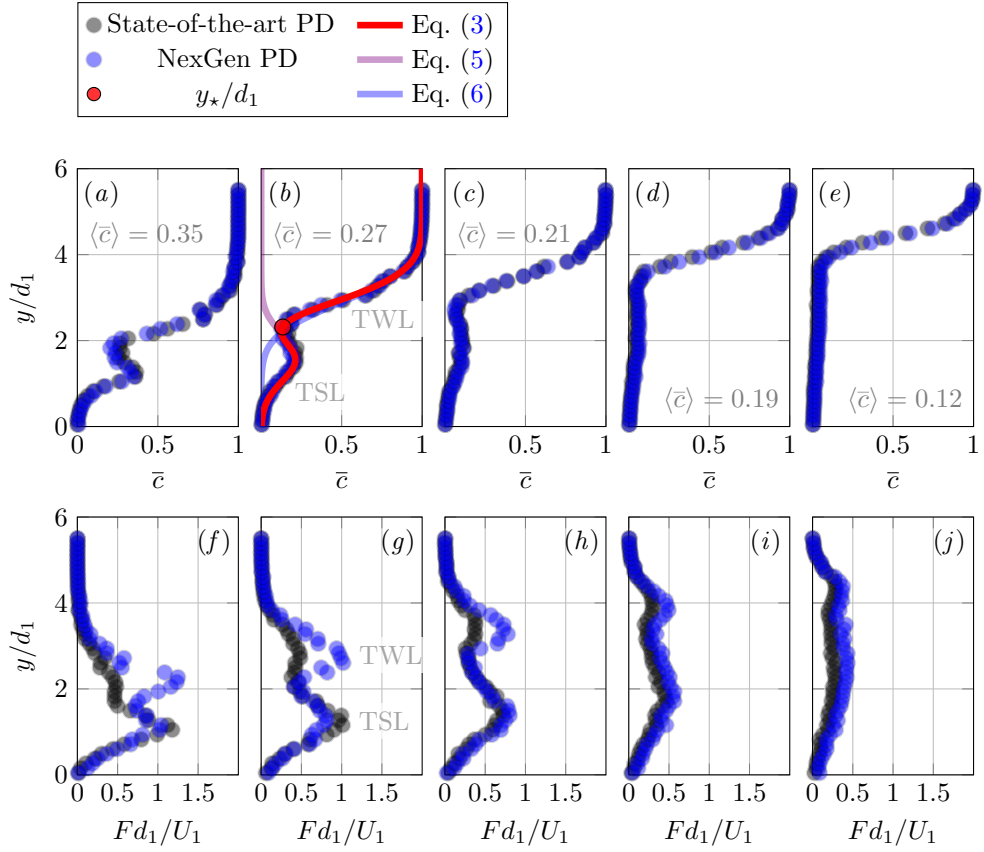}
\caption{Streamwise evolution of air concentrations and interface frequencies along the hydraulic jump; note that the mean air concentration $\langle \overline{c} \rangle$ is included for completeness: (\textit{a,f}) $(x-x_\text{toe})/d_1 = 3.1$; (\textit{b,g}) $(x-x_\text{toe})/d_1 =5.9$; $\mathcal{D} = 0.065$; $y_\star/d_1 = 2.32$; $\sigma_\star/d_1 = 0.45$; $\overline{c}_\text{max} = 0.22$; $y_{\overline{c}_\text{max}}/d_1 = 1.60$; (\textit{c,h}) $(x-x_\text{toe})/d_1 =8.3$;  (\textit{d,i}), $(x-x_\text{toe})/d_1 = 10.9$; (\textit{e,j}) $(x-x_\text{toe})/d_1 = 13.4$.}
\label{fig:airconcentration} 
\end{figure*}

Time averaged air concentrations $\overline{c}$ were evaluated from intrusive phase-detection measurements using Eq. (\ref{eq:timeaverage}). Here, Figs. \ref{fig:airconcentration}\textit{a} to \textit{e} show the streamwise evolution of air concentration profiles, together with corresponding interface frequencies (Figs. \ref{fig:airconcentration}\textit{f} to \textit{j}). 
Two distinct flow regions can be differentiated, comprising a Turbulent Shear Layer (TSL) and a Turbulent Wavy Layer (TWL, recirculation zone, jump roller), compare Fig. \ref{fig:experimentalsetup}\textit{a} and Figs. \ref{fig:airconcentration}\textit{b,g}, where the air concentration of the TWL mainly consists of entrapped air, i.e., air transported between wave peaks and throughs \citep{wilhelms2005bubbles,Kramer2024}. 

Adopting the two-state convolution approach from \cite{Kramer2023}, a theoretical solution for the air concentration distribution of the hydraulic jump is deduced herein, which comprises the air concentrations of the TSL ($\overline{c}_\text{TSL}$) and the TWL ($\overline{c}_\text{TWL}$) 
\begin{equation}
\overline{c}  = \overline{c}_\text{TSL} (1- \Gamma) + \overline{c}_\text{TWL} \Gamma,
\label{eq:twostate}
\end{equation}
with
\begin{equation}
\Gamma = \frac{1}{2} \left( 1+\erf \left(\frac{y - y_\star }{ \sqrt{2} \sigma_\star}  \right)  \right),
\label{eq:gaussianerr}
\end{equation}
where $\erf$ is the Gaussian error function, $y_\star$ corresponds to a time-averaged location of the interface/transition between the TSL and the TWL, and $\sigma_\star$ describes its standard deviation. The air concentration of the TSL is assumed to follow a solution of the advection-diffusion equation for air in water \citep{Chanson1996,Wang2014}
\begin{equation}
\overline{c}_\text{TSL} = \overline{c}_\text{max} \exp \left(-\frac{1}{4 \mathcal{D}} \frac{\left(\frac{y-y_{\overline{c}_\text{max}}}{d_1} \right)^2}{\left(\frac{x - x_\text{toe}}{d_1}\right)}   \right),
\label{eq:airconcTSL}
\end{equation}
while the expression for the air concentration of the TWL is written as \citep{VALERO201666,doi:10.1061/JHEND8.HYENG-13230,Kramer2023}
\begin{equation}
\overline{c}_{\text{TWL}} = \frac{1}{2} \, \left( 1 + \erf  \left( \frac{y - y_{50_\text{TWL}}}{ \sqrt{2} \, \mathcal{H} } \right) \right),
\label{eq:airconcTWL}
\end{equation}
where $\overline{c}_{\text{max}}$ is the peak air concentration of TSL, $y_{\overline{c}_{\text{max}}}$ its corresponding elevation, $\mathcal{D}$ is a dimensionless diffusivity, $y_{50_\text{TWL}}$ is the mixture flow depth where the free-surface air concentration is $\overline{c}_\text{TWL} = 0.5$, and $\mathcal{H}$ is a characteristic length-scale that describes the thickness/height of the TWL \citep{Kramer2023,Kramer2024}. As shown in Fig. \ref{fig:airconcentration}\textit{b}, the two-state convolution approach provided a novel physically-based solution for the air concentration distribution in hydraulic jumps. Note that associated dimensionless physical parameters $\mathcal{D}$, $y_\star/d_1$, $\sigma_\star/d_1$, $\overline{c}_\text{max}$, and $y_{\overline{c}_\text{max}}/d_1$ are listed in the caption of Fig. \ref{fig:airconcentration}\textit{b}, but a more detailed analysis is beyond the scope of the present work and is subject to future work. 

Measured interface frequencies (Figs. \ref{fig:airconcentration}\textit{f} to \textit{j}) exhibited two characteristic peaks, which are similarly associated with the TSL and the TWL. As the high velocity approach flow is entering into a body of slower water, the impingement point (or jump toe) acts as local point source for air bubbles, reflecting air concentration and interface frequency of the TSL. At the same time, the free-surface is highly deformed, which is due to the production of vorticity and turbulence, thereby exhibiting complex air-water surface features \citep{Wüthrich_Shi_Chanson_2021}. These wavy-like free-surface features, together with some entrained air, constitute the air concentration and interface frequency of the TWL. Related to the comparative analysis of the two phase-detection probes, an excellent agreement between the  state-of-the-art PD (black circles) and the NexGen PD (blue circles) is observed for the complete air concentration profiles and for interface frequencies of the TSL (Fig. \ref{fig:airconcentration}). However, due to reasons outlined in the last section, the NexGen PD was able to detect more interfaces within the TWL, which was most evident closest to the location of the jump toe (Figs. \ref{fig:airconcentration}\textit{f,g}).   

\begin{figure*}[h!]
\centering
\input{Figures/fig7.tex}
\caption{Probability density function estimates of air chord times $t_\text{ch,a}$ at $(x-x_\text{toe})/d_1 = 3.1$ for three selected elevations, including: (\textit{a}) Turbulent shear layer at $y/d_1 = 0.49$; (\textit{b}) Turbulent shear layer at $y/d_1 = 1.05$;  (\textit{c}) Turbulent wavy layer at $y/d_1 = 2.94$.}
\label{fig:chord}
\vspace{2cm}
\end{figure*}
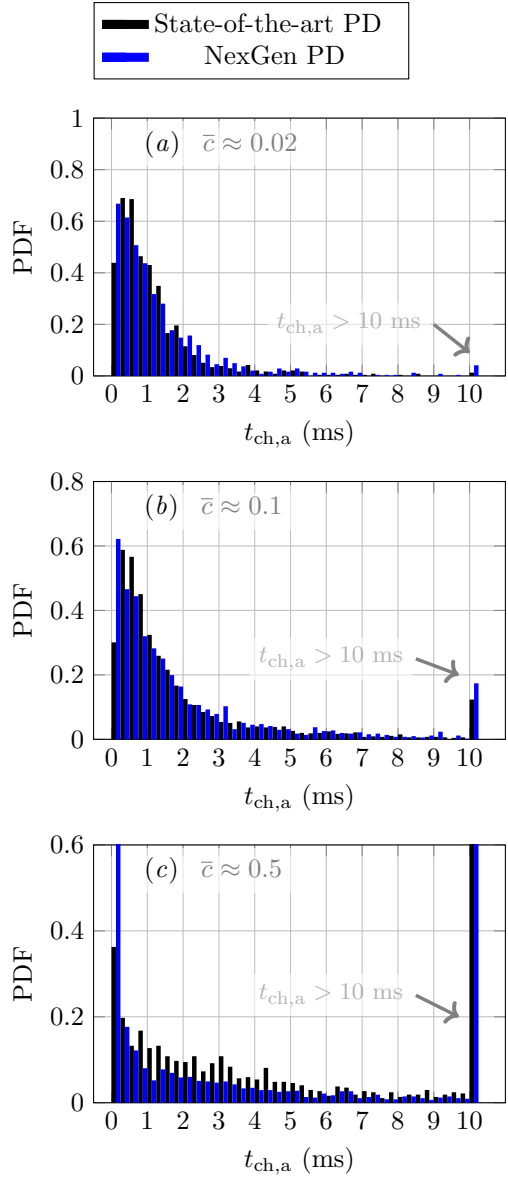

Other parameters of interest in air-water flow experiments are bubble size (or droplet size) distributions, which are herein presented in form of air chord times at the first measurement location $(x-x_\text{toe})/d_1 = 5.9$, comprising measurements in the TSL and the TWL (Fig. \ref{fig:chord}). The estimates of the probability density functions were computed using bin widths of 0.25 ms from 0 ms to 10 ms, i.e., the first bin ranges from 0 ms to 0.25 ms, the second from 0.25 ms to 0.5 ms, and so on. In order to display the results from both probes, the bin widths were reduced to a width of 0.125 ms for plotting purposes, and shifted by 0.0625 ms (NexGen PD, blue bars) and -0.0625 ms (state-of-the-art PD, black bars). Note that cord times larger than 10 ms were collected in the last bin. The data of the first two measurements locations in the TSL exhibited Gamma-type distributions (Fig. \ref{fig:chord}\textit{a,b}), while the distribution of air chord times within the TWL was flatter, which is likely a result from the measurement of entrapped air rather than entrained air. It is further noted that the mean air chord time, averaged over both probes, increased from 1.31 ms at $y/d_1 = 0.49$, over 2.55 ms at $y/d_1 = 1.05$, to 15.34 ms at $y/d_1 = 2.94$, which is a reflection of the different flow regions. The comparison of the chord time data for the two probes showed a good agreement in the TSL (Fig. \ref{fig:chord}\textit{a,b}), while, in line with previous discussions, some differences were found in the TWL  (Fig. \ref{fig:chord}\textit{c}). Overall,  Fig. \ref{fig:chord} suggests very small effects of probe type on measured chord times in the current experiments.

\subsection{Interfacial velocities and turbulence}
\begin{figure*}[h!]
\centering
\input{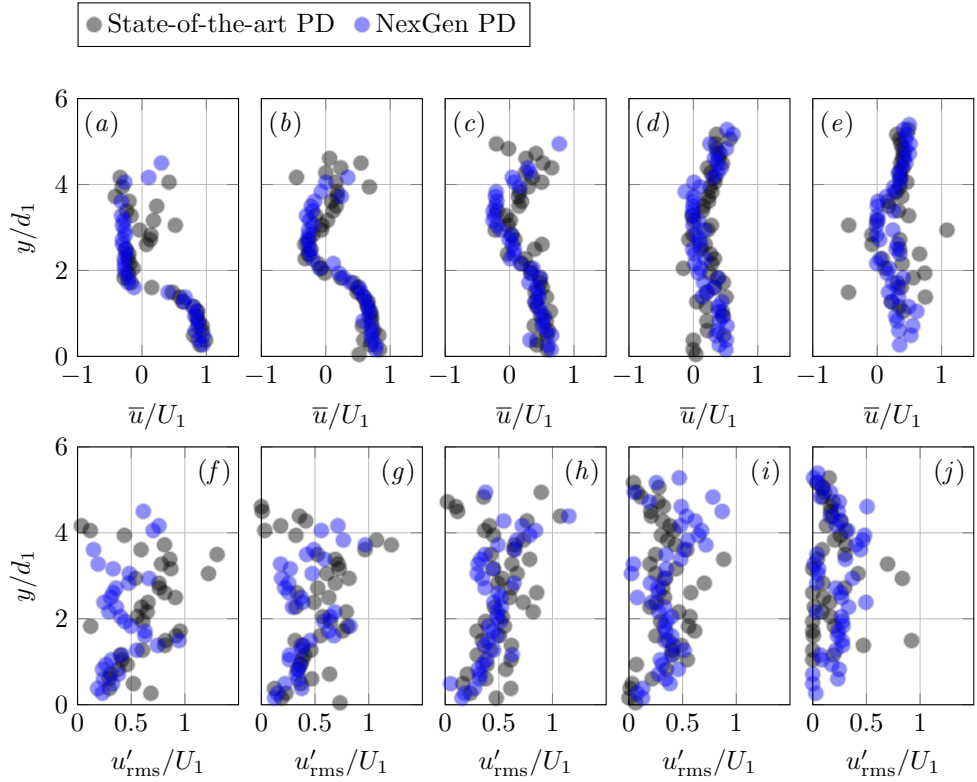}
\caption{Streamwise evolution of interfacial velocities and their fluctuations along the hydraulic jump: (\textit{a,f}) $(x-x_\text{toe})/d_1 = 3.1$; (\textit{b,g}) $(x-x_\text{toe})/d_1 =5.9$; (\textit{c,h}), $(x-x_\text{toe})/d_1 =8.3$; (\textit{d,i}), $(x-x_\text{toe})/d_1 = 10.9$; (\textit{e,j}) $(x-x_\text{toe})/d_1 = 13.4$.}
\label{fig:velocities}
\end{figure*}

The streamwise evolution of interfacial velocity distributions of the hydraulic jump is shown in Figs. \ref{fig:velocities}\textit{a} to \textit{d}. Note that the streamwise velocity was normalised with the inflow velocity $U_1 = q/d_1$. Close to the jump toe, the interfacial velocity profiles displayed some similarity to a classical wall jet \citep{doi:10.1061/JYCEAJ.0001299}, while a more uniform velocity distribution was observed farther downstream, i.e., Fig. \ref{fig:velocities}\textit{d}. The velocity above the channel bed increased in bed-normal direction, exhibiting a boundary layer type flow, which was followed by a TSL, characterised by a decrease in streamwise velocity. Above the shear layer, a recirculation zone with negative flow velocities occurred, implying flow reversal, a wavy free-surface (TWL), and the formation of the jump roller. It is noted that these different flow momentum layers are causing turbulent mass transfer, and therefore, they are intrinsically linked to the air concentration distribution. At any given location, the interfacial velocity distributions estimated using the state-of-the-art PD (black circles) and the NexGen CP (blue circles) showed some good agreement. 

For completeness, the root-mean-square of velocity fluctuations $u'_\text{rms}$ was evaluated for the five recorded profiles, shown in Figs. \ref{fig:velocities}\textit{f} to \textit{j}. Close to the jump toe, the maximum fluctuation occurred near the air-water interface (Fig. \ref{fig:velocities}\textit{f}), and this peak shifted further up as one moved downstream, eventually decaying towards the end section of the hydraulic jump. Overall, the shape of recorded velocity fluctuations showed some resemblance with direct numerical simulations carried out by \cite{Mortazavi2016}, while some scatter within the recirculation region is acknowledged. As discussed in \cite{Kramer20Practicescomment}, the estimation of interfacial velocities in highly three-dimensional flow regions, such as the recirculation region in hydraulic jumps, violates the assumption of the probe tips being aligned with flow streamlines. As such, caution must prevail as a velocity bias due to probe misalignment has likely been introduced. Nonetheless, the overall good agreement between the state-of-the-art PD and the NexGen PD data is (again) apparent.

\section{Discussion}
\label{sec:discussion}

\subsection{Quantitative comparison of key air-water flow parameters}
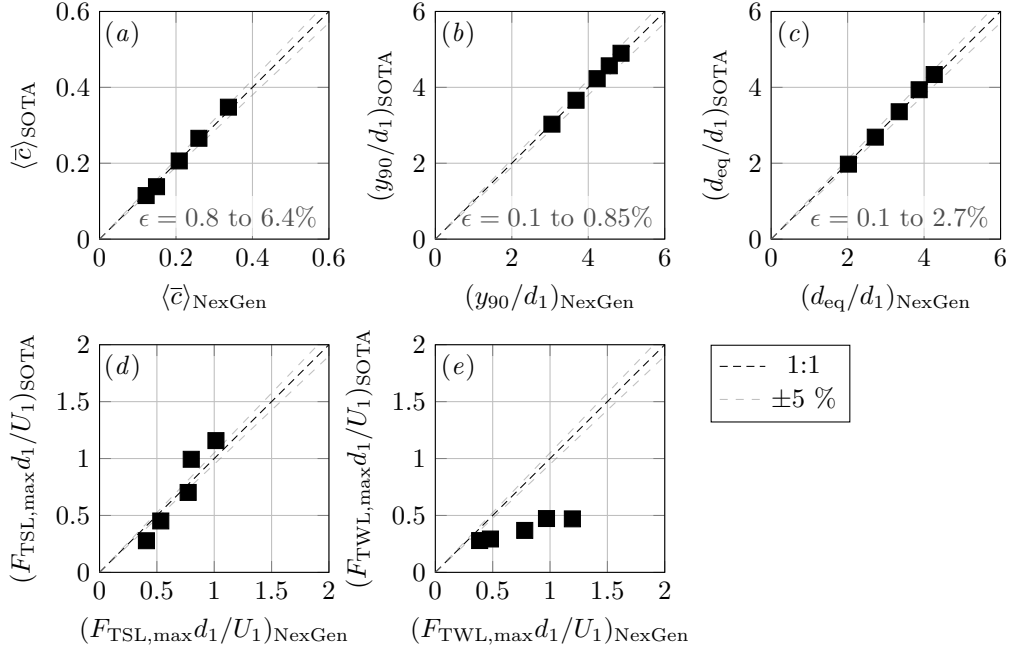
\begin{figure*}[h!]
\centering
\input{Figures/fig8.tex}
\caption{Comparison of statistical design parameters evaluated using a state-of-the-art PD and the NexGen PD; SOTA = state-of-the-art: (\textit{a}) Mean air concentration $\langle \overline{c} \rangle$; (\textit{b}) Mixture flow depth $y_{90}$; (\textit{c}) Equivalent clear water flow depth $d_\text{eq}$; (\textit{d}) Maximum interface frequency $F_\text{max,TSL}$ of the TSL; (\textit{e}) Maximum interface frequency $F_\text{max,TWL}$ of the TWL.}
\label{comparison}
\end{figure*}

In the previous $\S$ \ref{sec:results}, voltage signals, air concentrations, interface frequency distributions, and chord times, as well as interfacial velocity profiles and their fluctuations, were measured using the NexGen PD and a state-of-the-art PD, showing an overall good agreement. As this previous comparison was mostly of qualitative nature, a more quantitative approach is selected by computing the percentage deviation of key air-water flow parameters, which are important for hydraulic structures design. These include the mean air concentration $\langle \overline{c} \rangle = \frac{1}{y_{90}}\int_{y=0}^{y_{90}} \overline{c} \, \text{d}y$, the mixture flow depth $y_{90}$, the equivalent clear water flow depth $d_\text{eq} = \int_{y=0}^{y_{90}}  (1-\overline{c}) \, \text{d}y$, and the maximum interface frequencies of the TSL and the TWL ($F_\text{TSL,max}$ and $F_\text{TWL,max}$), respectively. The percentage deviation was calculated, for example for the mean air concentration, as follows
\begin{equation}
\epsilon = \frac{\vert \langle \overline{c} \rangle_\text{NexGen} - \langle \overline{c} \rangle_\text{SOTA} \vert}{ 0.5 \left(\langle \overline{c} \rangle_\text{NexGen} + \langle \overline{c} \rangle_\text{SOTA}\right)} \, 100,
\end{equation}
where SOTA stands for state-of-the-art. The respective ranges of $\epsilon$ are embedded in Fig. \ref{comparison}, where a comparison of air-water flow design parameters is presented. 

The largest differences were found in relation to interface frequencies  (sometimes referred to as \q{bubble count rates}), see Figs. \ref{comparison}\textit{d,e}, which is expected, as a high dependency of $F$ on the shape and size of the tip has been reported previously \citep{Chanson2016,FELDER201788,SHI2023102479}. 
Interestingly, the NexGen PD was able to systematically detect more interfaces than the state-of-the-art PD, which is despite the fact that the tip has not been sharpened. It is believed that this was due to the relatively small size of both electrodes as well as due to the different electronic system, and as such, the NexGen PD holds high potential for future developments/modifications, including tip layout, dimensions, and shape, which is additionally facilitated by the detachable probe head. Overall, comparing the present small deviation in other key air-water flow parameters ($\lessapprox 5\%$; Figs. \ref{comparison}\textit{a} to \textit{c})  with published measurement uncertainties of dual-tip phase-detection intrusive probes \cite[Table 1]{PAGLIARA2024104660}, it can be concluded that NexGen PD is well suited for air-water flow experiments.

\subsection{Schmitt-Trigger and hysteresis}
\label{sec:hysteresis}
Here, the previously outlined considerations related to the electronic box/air bubble detector A25240 are briefly re-iterated. The first version of the air bubble detector A25240 was built by the New Zealand Institute for Industrial Research and Development (NZIIRD) in the 1980ies \citep{Chanson1988}, and the circuit contains a Schmitt-Trigger \citep{Cummings1996}, which exhibits a hysteresis using an upper and a lower voltage threshold, aiming to remove some noise and converting the analogue signal into a digital waveform. It is anticipated that the setting of the two thresholds has some non-negligible influence on the air-water flow results, similar to the single-threshold level typically used in signal post-processing \citep{Felder2015,SHI2023102479}. However, different voltage levels for upper and lower thresholds have not been reported/investigated in the past, which is believed to result from the fact that the adjustment of the Schmitt-Trigger requires some detailed knowledge of the circuit and associated electronics. In this context, it is anticipated that the newly developed electronics circuit, which delivers the analogue raw signals of the probe tips, allows for a better signal interpretation and processing. 

\subsection{Potential limitations: Probe tip deterioration}

During the present experiments, some electrochemical corrosion of the gold electrodes of the NexGen PD was observed, which has been similarly reported by previous researchers for other materials. For example, NZIIRD noted that a needle sensor with an inner electrode made of Copper ($\Phi_\text{in} = 0.2$ mm) only had a life time of one day, despite using a voltage switching timer at 0.5 Hz, therefore recommending the the manufacturing of a platinum tip probe
\citep{Cummings1996}. Since then, platinum has been used as preferred material for inner electrodes of phase-detection conductivity probes, compare Table \ref{table:intro}.

\begin{figure*}[h!]
\centering
\input{Figures/fig9.tex}
\caption{Normalised voltage span of the NexGen probe tips throughout the additional open channel flow experiment; note that the signal was segmented into chunks of $T = 100$ s for this analysis.}
\label{fig:limitations}
\end{figure*}
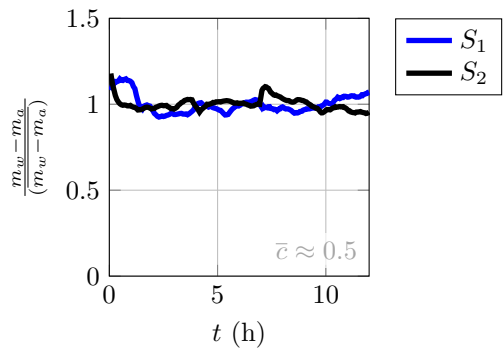

In order to document the deterioration of the probe tips, an additional experiment was performed, where the NexGen probe was placed close to the free-surface ($\overline{c} \approx 0.5$) of an open channel flow for an extended duration of $t = 12$ h. For the subsequent analysis, the complete signal was segmented into chunks of $T = 100$ s, and the corresponding time evolution of the voltage span ($m_w - m_a$) of both tips, normalised with their time averaged averaged value ($\overline{m_w - m_a}$), is shown in Fig. \ref{fig:limitations}. It is seen that the voltage span remained relatively constant for both tips, which implies that no major deterioration occurred throughout the twelve hours of measurements, while it is acknowledged that the life span of a probe tip is ideally in the order of hundreds or thousands of hours. In order to extend the life span of the probe tips, the electronic box was modified such that the emitted voltage across the sensor's electrodes was reduced from 5 V to 0.8 V. Since the modification, no more corrosion issues have been noticed.

\section{Conclusion}
In the present study, the development of the next generation of dual-tip phase-detection conductivity probes for air-water flow experiments is presented. The key novelties of the NexGen phase-detection probes comprise a circuit board design of the two sensor tips, together with a detachable sensor head. These features are anticipated to improve quality and repeatability of air-water flow measurements, simply because the circuit boards can be produced to a high standard with low manufacturing tolerances, and because the detachable sensor head allows for a quick exchange of the sensor's most fragile part, i.e., the probe tips, in case of damage or deterioration. 

The NexGen phase-detection probe was validated against a state-of-the-art system within the highly-aerated flow of a hydraulic jump with inflow Froude number of $Fr_1 = 3.61$. Measured air concentration distributions, interface frequencies, interfacial velocities, their fluctuations, as well as other key air-water flow parameters were in excellent agreement between the two systems, demonstrating the suitability of the NexGen PD for air-water flow experiments. It is noted that different electronic circuits were used throughout this work, and that the state-of-the-art circuit exhibits some hysteresis effects. As such, identical results were not expected, but a slight adjustment of the single threshold value of the NexGen PD led to a good agreement in the comparative analysis. In this context, the need for future systematic studies on the effects of different electronic circuit layouts on computed air-water flow properties has been raised. 

Overall, it is hoped that the current findings will spark new developments in the design of multiphase flow measurement instrumentation for hydraulic engineering applications, thereby enabling broader research community access to these delicate measuring devices. 

\section*{Data Availability Statement}
Some or all data, models, or code that support the findings of this study are available from the corresponding author upon reasonable request.

\section*{Acknowledgments}
The author would like to thank Edward Sharrer, Brendan Wallace, David Sharp, Matthew Barret and Troy Huber for their support in developing and manufacturing the next generation of dual-tip phase-detection probes. Dr Kapil Chauhan (University of Sydney) and Dr Elena Pummer (NTNU Trondheim) are thanked for providing a state-of-the-art phase-detection probe system, used for comparative analyses, and for sharing their images. 

\section*{Declaration of competing interest}
The authors declares that he has no known competing financial interests or personal relationships hat could have appeared to influence the work reported in this paper.

\section*{Notation} \label{sec:notation}
\emph{The following symbols are used in this paper:}
\vspace{-0.5cm}
\begin{tabbing}
\hspace*{0cm}\=\hspace*{2cm}\=\hspace*{11.2cm}\=\kill\\
\>$c$  \> instantaneous volumetric air concentration  \> (-)\\
\>$\overline{c}$  \> air concentration (time-averaged) \> (-)\\
\>$\overline{c}_\text{max}$  \> maximum air concentration of the TSL \> (-)\\
\>$\langle \overline{c} \rangle$  \> depth-averaged air concentration $\langle \overline{c} \rangle = \frac{1}{y_{90}}\int_{y=0}^{y_{90}} \overline{c} \, \text{d}y$ \> (-)\\
\>$d_{1}$  \> supercritical conjugate flow depth \> (m)\\
\>$d_{2}$  \> subcritical conjugate flow depth \> (m)\\
\>$d_\text{eq}$  \> equivalent clear water flow depth $d_\text{eq} = \int_{y=0}^{y_{90}}  (1-\overline{c}) \, \text{d}y$ \> (m)\\
\>$\mathcal{D}$  \> dimensionless turbulent diffusivity \> (-)\\
\>$F$  \> interface frequency  \> (s$^{-1}$)\\
\>$Fr_1$  \> inflow Froude number $Fr_1 = U_1/\sqrt{g d_1}$ \> (-)\\
\>$g$  \> gravitational acceleration \> (m$^2$  s$^{-1}$)\\
\>$h_\text{gate}$  \> height of downstream sluice gate \> (m)\\
\>$\mathcal{H}$  \> length-scale of the TWL \> (m)\\
\>$L_\text{jump}$  \> length of hydraulic jump \> (m)\\
\>$m$  \> mode of probability density function \> (V)\\
\>$n_p$  \> number of interface pairs/particles used for AWCC analysis \> (-)\\
\>$P$  \> height of sharp crested weir \> (m)\\
\>$q$  \> specific water discharge \> (m$^2$ s$^{-1}$)\\
\>$S$  \> raw signal of phase-detection probe tip \> (V)\\
\>$t$  \> time \> (s)\\
\>$t_\text{ch}$  \> chord time \> (s)\\
\>$T$  \> sampling duration \> (s)\\
\>$\mathcal{T}$  \> single threshold for signal binarization  \> (-)\\
\>$\overline{u}$  \> streamwise air-water velocity (time-averaged) \> (m s$^{-1}$)\\
\>$\langle \overline{u} \rangle$  \> depth-averaged air-water velocity $\langle \overline{u} \rangle = \frac{1}{y_{90}}\int_{y=0}^{y_{90}} \overline{u} \, \text{d}y$  \> (m s$^{-1}$)\\ 
\>$u'$  \> fluctuating part of the streamwise velocity \> (m s$^{-1}$)\\
\>$U_1$  \> approach flow velocity (averaged in time and space) \> (m s$^{-1}$)\\
\>$x$  \> streamwise coordinate \> (m)\\ 
\>$x_\text{toe}$  \> jump toe position \> (m)\\ 
\>$y$  \> coordinate normal to the channel bed \> (m)\\ 
\>$y_{\overline{c}_\text{max}}$  \> location of peak air concentration of the TSL \> (m)\\ 
\>$y_\star$  \> time-averaged interface location \> (m)\\ 
\>$y_{50}$  \> mixture flow depth where $\overline{c} = 0.5$ \> (m)\\
\>$y_{90}$  \> mixture flow depth where $\overline{c} = 0.9$ \> (m)\\
\>$\Gamma$  \> cumulative distribution function of interface position \> (-)\\ 
\>$\Delta x$  \> probe-wise tip separation distance \> (m)\\ 
\>$\Delta z$  \> probe-normal tip separation distance \> (m)\\
\>$\epsilon$  \> percentage deviation \> (\%)\\
\>$\sigma_\star$  \> standard deviation of interface position \> (m)\\
\>$\Phi$  \> tip or electrode diameter \> (m)\\ 
\>$\Phi_\text{in}$  \> diameter of inner electrode \> (m)\\ 
\>$\Phi_\text{out}$  \> diameter of outer electrode \> (m)\\
\end{tabbing}
\textbf{Indices and operators}
\vspace{-0.5cm}
\begin{tabbing}
\hspace*{0cm}\=\hspace*{2cm}\=\hspace*{9.2cm}\=\kill\\
\>$1$  \> leading tip \> \\
\>$a$  \> air \> \\
\>AWCC  \> adaptive window cross correlation \> \\
\> CBR  \> Canberra \> \\
\> NexGen  \> next generation \> \\
\> NZIIRD  \> New Zealand Institute for Industrial Research and Development \> \\
\> max  \> maximum \> \\
\> PCB  \> printed circuit board \> \\
\> PD  \> phase-detection intrusive probe \> \\
\> rms  \> root-mean-square \> \\
\> SOTA  \> state-of-the-art \> \\
\> TSL  \> turbulent shear layer \> \\
\> TWL  \> turbulent wavy layer \> \\
\>$w$  \> water \> \\
\>$\overline{\phantom{C}}$  \> time averaging \> \\
\>$\langle \phantom{C} \rangle $ \> spatial averaging \> \\
\end{tabbing}

\bibliography{sn-bibliography}

\end{document}

%% file: Figures/tab1.tex
\begin{footnotesize}
\caption{Key developments of phase-detection intrusive conductivity probes for air-water flow measurements in hydraulic engineering applications; $\Phi_\text{in} = $ inner electrode diameter; $\Phi_\text{out} = $ outer electrode diameter;  $\Delta x =$ probe-wise tip separation distance; $\Delta z =$ probe-normal tip separation distance; CBR = Canberra}
\begin{tabular}{p{5 cm} p{4.3 cm} p{4.5 cm} p{4.5 cm} c }
\toprule
Reference &  Institution & Type of air-water flow & Comment \\
\midrule
\cite{Lamb1950}$^1$, \cite{Straub1958}$^1$, \cite{killen1968}$^{1,2}$  & Univ. of Minnesota, US &Smooth laboratory spillway & First development of phase-detection conductivity probes\\
\cite{Keller1972}$^3$, \cite{Cain1978}$^{4,5}$, \cite{Cain1981}$^{4,5}$ & Univ. of Canterbury, NZ & Smooth laboratory and prototype (Aviemore Dam, NZ) spillways & Design adapted from \cite{Lamb1950}\\
\cite{Volkart1978}$^6$ &  ETH Zurich, CH & Smooth laboratory spillway & Developed at ETH Zurich \\
\cite{Tan1984}, \cite{Low1986}, \cite{Chanson1988}$^7$ & Univ. of Canterbury, NZ  & Laboratory spillway aerator & Design adapted from \cite{Cain1981}\\ 
\cite{Frizell1994}$^8$, \cite{Matos1997}$^8$ & Bureau of Reclamation, US  & Laboratory stepped spillway & Design adapted from Univ. of Canterbury probes \\ 
\cite{Chanson1995}$^9$, \cite{Cummings1997}$^9$, \cite{Chanson2016} & Univ. of Queensland, AU & Laboratory hydraulic jump, plunging jet, smooth and stepped spillways  & Design adapted from \cite{Chanson1988}\\ 
\cite{Thorwarth2008}$^{10}$ & RWTH Aachen, GER & Laboratory stepped spillway & Design adapted from \cite{Chanson1988}\\ 
\cite{Felder2016}$^{11}$ &   Univ. of New South Wales, AU  &  Laboratory hydraulic jump &  Developed at UNSW\\
\cite{Tang2022}$^{12}$ &  Sichuan Univ., CN & Prototype hydraulic jump & Design adapted from \cite{Chanson1995}\\
Present study$^{13}$ &  Univ. of New South Wales CBR, AU & Laboratory hydraulic jump & Developed at UNSW Canberra\\
\bottomrule
\end{tabular}
\end{footnotesize}
\begin{flushleft}  
\footnotesize{$^1$ Conductivity probe with two different electrodes, separated by 6.35 mm\\
$^2$ Additionally deployed a so-called dipping probe to measure free-surface roughness\\
$^3$ First design of single-tip probe with $\Phi_\text{in} = 6.35$ mm\\
$^4$ Combined stagnation pressure and air concentration probe; 12 mm separation distance between the two electrodes\\
$^5$ First design of dual-tip correlation probe for velocity measurements; stainless steel electrodes ($\Phi_\text{in} = 1 $ mm); $\Delta x = 101.6$ mm; $\Delta z = 0$ mm and 6.4 mm \\
$^6$ Conductivity probe with two different electrodes ($\Phi = 0.5$ mm), separated by 1.5 mm\\
$^7$ Dual-tip needle conductivity probe with platinum wire ($\Phi_\text{in} = 0.2$ mm) and annular needle ($\Phi_\text{out} = 0.8$ mm);\\
$\phantom{^7}$ $\Delta x = 10$ mm; $\Delta z = 0$ mm;  Electronic system/Air bubble detector A25240\\
$^8$ Single-tip needle conductivity probe with platinum wire ($\Phi_\text{in} = 0.2$ mm) and stainless steel tube ($\Phi_\text{out} = 1.59$ mm)\\
$^9$ Dual-tip needle conductivity probe with  platinum wire ($\Phi_\text{in} = 0.025$ mm) and annular needle ($\Phi_\text{out} = 0.2$ mm);\\ 
$\phantom{^9}$ $\Delta x = 8$ mm; $\Delta z = 1.33$ mm; Electronic system/Air bubble detector A25240\\
$^{10}$ Dual-tip needle conductivity probe platinum–iridium electrode ($\Phi_\text{in} = 0.13$ mm) and steel tube ($\Phi_\text{out} = 0.4$ mm); $\Delta x = 5$ mm; $\Delta z = 0.7$ mm \\
$^{11}$ Dual-tip needle conductivity probe with  annular needle ($\Phi_\text{out} = 0.5$ mm) and platinum wire ($\Phi_\text{in} = 0.125$ mm); \\
$\phantom{^{11}}$ $\Delta x = 7.9$ mm; $\Delta z = 1$ mm; Electronic system/Air bubble detector modified from A25240\\ 
$^{12}$ Thick-tip needle conductivity probe with stainless steel tube ($\Phi_\text{out} = 3$ mm) and tungsten wire; ($\Phi_\text{in} = 1$ mm) $\Delta x = 10$ mm; $\Delta z = 2 $ mm \\ 
$^{13}$ NexGen dual-tip phase-detection conductivity probe with gold electrodes; electrode diameters $\Phi = 0.2$ mm; circuit board thickness = 0.2 mm; \\
$\phantom{^{13}}$ $\Delta x = 5$ mm; $\Delta z = 1 $ mm
}\end{flushleft} 

%% file: Figures/tab2.tex
\begin{center}
\begin{table}[h!]
\caption{Experimental flow conditions}
\begin{tabular}{c c c c c c c c}
\toprule
$q$ & $P$  & $h_\text{gate}/P$ & $Fr_1$ & $x_\text{toe}/P$ & $L_\text{jump}/d_2$ & $d_2/d_1$  \\
 (m$^2$/s) &   (m) & (-) & (-)  & (-)  & (-) & (-) \\
\midrule
 0.11 &  0.4 & 0.23 & 3.61 & 4.16 & 8.44 & 4.8\\
\bottomrule
\end{tabular}
\label{table:conditions}
\end{table}
\end{center}

%% file: Figures/fig4.tex
\begin{tikzpicture}

\begin{groupplot}[
     group style = {group size = 3 by 1,horizontal sep=1.3cm,vertical sep=1.4cm},
    width = 1\textwidth]

\nextgroupplot  [width=4.6cm, height=4.6cm,
yshift=-5.3cm,
    xlabel={PDF},
    ylabel={$(S_1 - m_a)/(m_w - m_a)$},
    grid=both,
    legend style={cells={align=left},anchor = south west,at={(0,1.2)},font=\normalsize},
   clip mode=individual,
    legend columns=2,
   xmin=0,
   xmax=35,
  ymin=-0.5,
  ymax=1.5,
   ]

\node[align=center] at (rel axis cs:0.1,0.9) {(\textit{b})};

\addplot+[xbar,mark=none,bar width=0.035, draw =black, fill = black, fill opacity = 0.8, draw opacity = 0] 
table {
2.45821042281219e-05	-0.371465279037928
4.91642084562438e-05	-0.356724394873747
4.91642084562441e-05	-0.341983510709566
0.000122910521140609	-0.327242626545386
0.000294985250737463	-0.312501742381205
0.00145034414945919	-0.297760858217024
0.00319567354965585	-0.283019974052844
0.00621927236971488	-0.268279089888663
0.0149459193706981	-0.253538205724482
0.0183136676499508	-0.238797321560301
0.0158554572271386	-0.224056437396121
0.0206981317600786	-0.209315553231940
0.0310226155358898	-0.194574669067759
0.0394788593903638	-0.179833784903579
0.0489921337266473	-0.165092900739398
0.0558505408062929	-0.150352016575217
0.0724680432645034	-0.135611132411037
0.0794739429695186	-0.120870248246856
0.0854965585054080	-0.106129364082675
0.100147492625369	-0.0913884799184945
0.107890855457227	-0.0766475957543137
0.109021632251721	-0.0619067115901330
0.107571288102261	-0.0471658274259523
0.113864306784661	-0.0324249432617716
0.122787610619469	-0.0176840590975909
0.131858407079647	-0.00294317493341013
0.128638151425762	0.0117977092307705
0.132104228121926	0.0265385933949511
0.119321533923305	0.0412794775591320
0.107694198623402	0.0560203617233126
0.0943461160275319	0.0707612458874933
0.0754424778761061	0.0855021300516740
0.0628564405113077	0.100243014215855
0.0512045231071779	0.114983898380035
0.0431415929203545	0.129724782544216
0.0402900688298918	0.144465666708397
0.0350786627335300	0.159206550872577
0.0314405113077679	0.173947435036758
0.0309242871189774	0.188688319200939
0.0275073746312684	0.203429203365120
0.0283431661750245	0.218170087529300
0.0250000000000000	0.232910971693481
0.0248279252704031	0.247651855857662
0.0220993117010816	0.262392740021842
0.0226892822025565	0.277133624186023
0.0218043264503441	0.291874508350204
0.0201573254670600	0.306615392514385
0.0201081612586037	0.321356276678565
0.0180924287118977	0.336097160842746
0.0184857423795477	0.350838045006927
0.0175024582104230	0.365578929171107
0.0174041297935103	0.380319813335288
0.0182153392330383	0.395060697499469
0.0157571288102261	0.409801581663649
0.0172566371681416	0.424542465827830
0.0157325467059980	0.439283349992011
0.0153392330383483	0.454024234156192
0.0159537856440511	0.468765118320372
0.0149950835791544	0.483506002484553
0.0167158308751229	0.498246886648734
0.0148230088495575	0.512987770812914
0.0153146509341199	0.527728654977095
0.0140363815142576	0.542469539141276
0.0147738446411013	0.557210423305456
0.0142330383480826	0.571951307469637
0.0145771878072763	0.586692191633818
0.0143067846607669	0.601433075797999
0.0145034414945919	0.616173959962179
0.0139134709931170	0.630914844126360
0.0142822025565390	0.645655728290541
0.0139626352015732	0.660396612454721
0.0121681415929203	0.675137496618902
0.0130039331366765	0.689878380783083
0.0135693215339233	0.704619264947264
0.0131268436578171	0.719360149111444
0.0123156342182891	0.734101033275625
0.0126843657817109	0.748841917439806
0.0135201573254670	0.763582801603986
0.0134218289085546	0.778323685768167
0.0135693215339235	0.793064569932348
0.0130285152409046	0.807805454096528
0.0128072763028515	0.822546338260709
0.0126597836774828	0.837287222424890
0.0128072763028515	0.852028106589071
0.0126843657817109	0.866768990753251
0.0132989183874139	0.881509874917432
0.0123647984267453	0.896250759081612
0.0135939036381514	0.910991643245794
0.0124877089478859	0.925732527409974
0.0124139626352016	0.940473411574155
0.0196165191740418	0.955214295738335
0.0681415929203521	0.969955179902516
0.332694198623411	0.984696064066697
25.1642576204516	0.999436948230877
0.697148475909555	1.01417783239506
0.121017699115041	1.02891871655924
0.0312438544739437	1.04365960072342
0.00442477876106183	1.05840048488760
0.00147492625368735	1.07314136905178
0.000639134709931153	1.08788225321596
};

\addplot+[xbar,mark=none,bar width=0.035, draw =black, fill = blue, fill opacity = 0.8, draw opacity = 0] 
table {
2.93427230046949e-05	-0.565140116948296
2.93427230046949e-05	-0.547764010149947
0	-0.530387903351598
2.93427230046949e-05	-0.513011796553249
0.000498826291079813	-0.495635689754899
0.000909624413145541	-0.478259582956550
0.000674882629107977	-0.460883476158201
0.000322769953051643	-0.443507369359852
0.000410798122065728	-0.426131262561503
0.000381455399061033	-0.408755155763154
0.000968309859154930	-0.391379048964804
0.00355046948356808	-0.374002942166455
0.00202464788732395	-0.356626835368106
0.00190727699530517	-0.339250728569757
0.00478286384976522	-0.321874621771408
0.00765845070422536	-0.304498514973058
0.00830399061032864	-0.287122408174709
0.0151701877934272	-0.269746301376360
0.0291079812206573	-0.252370194578011
0.0363849765258216	-0.234994087779662
0.0369424882629108	-0.217617980981313
0.0488556338028169	-0.200241874182964
0.0620011737089202	-0.182865767384614
0.0732394366197184	-0.165489660586265
0.0651408450704226	-0.148113553787916
0.0600352112676057	-0.130737446989567
0.0680164319248827	-0.113361340191218
0.0612676056338029	-0.0959852333928687
0.0595363849765259	-0.0786091265945195
0.0554870892018775	-0.0612330197961703
0.0527875586854460	-0.0438569129978211
0.0515845070422536	-0.0264808061994719
0.0524061032863850	-0.00910469940112275
0.0550176056338029	0.00827140739722640
0.0482981220657274	0.0256475141955756
0.0479753521126761	0.0430236209939250
0.0488849765258224	0.0603997277922739
0.0473884976525815	0.0777758345906228
0.0475645539906104	0.0951519413889722
0.0489436619718310	0.112528048187322
0.0485915492957747	0.129904154985670
0.0506748826291080	0.147280261784020
0.0479753521126761	0.164656368582369
0.0492957746478874	0.182032475380718
0.0512323943661972	0.199408582179067
0.0527582159624413	0.216784688977417
0.0513791079812207	0.234160795775765
0.0556924882629108	0.251536902574115
0.0553990610328630	0.268913009372464
0.0528462441314563	0.286289116170813
0.0529636150234734	0.303665222969162
0.0546948356807521	0.321041329767511
0.0531396713615016	0.338417436565860
0.0539612676056347	0.355793543364210
0.0557805164319241	0.373169650162559
0.0600645539906113	0.390545756960908
0.0603579812206564	0.407921863759257
0.0601819248826301	0.425297970557606
0.0624119718309850	0.442674077355955
0.0615316901408451	0.460050184154305
0.0644659624413146	0.477426290952654
0.0663732394366198	0.494802397751003
0.0662852112676057	0.512178504549352
0.0678110328638498	0.529554611347701
0.0723004694835681	0.546930718146051
0.0720657276995306	0.564306824944400
0.0739436619718310	0.581682931742749
0.0733274647887325	0.599059038541098
0.0756161971830986	0.616435145339447
0.0811619718309847	0.633811252137796
0.0806924882629121	0.651187358936146
0.0826291079812207	0.668563465734495
0.0877053990610329	0.685939572532844
0.0885563380281691	0.703315679331193
0.0895833333333320	0.720691786129542
0.0950117370892020	0.738067892927892
0.0995305164319249	0.755443999726241
0.104518779342723	0.772820106524590
0.105897887323944	0.790196213322939
0.114876760563380	0.807572320121288
0.122241784037559	0.824948426919637
0.135299295774648	0.842324533717987
0.146537558685446	0.859700640516335
0.166431924882629	0.877076747314685
0.211766431924883	0.894452854113034
0.263849765258216	0.911828960911383
0.367664319248827	0.929205067709732
0.543133802816902	0.946581174508082
1.02529342723005	0.963957281306430
3.95399061032864	0.981333388104780
14.6605340375587	0.998709494903129
8.33013497652583	1.01608560170148
1.16065140845071	1.03346170849983
0.134947183098592	1.05083781529818
0.0251173708920188	1.06821392209653
0.00469483568075118	1.08559002889487
0.000733568075117360	1.10296613569322
0.000352112676056338	1.12034224249157
5.86854460093897e-05	1.13771834928992
2.93427230046949e-05	1.15509445608827
};

 \node[align=center, fill=white, fill opacity=0.7, anchor=east] at (rel axis cs:0.99,0.1) {$ \overline{c}  \approx 0.1$};

 \node[align=center, fill=white, fill opacity=0.0, anchor=west,text opacity=1] at (27,1) {$m_w$};

\addplot+[dashed,mark=none, draw =black, draw opacity = 1] 
table {
0 0.5
35 0.5
};

\addplot+[dashed,mark=none, draw =blue, draw opacity = 1] 
table {
0 0.65
35 0.65
};

\nextgroupplot  [width=4.6cm, height=4.6cm,
    xlabel={PDF},
    grid=both,
    legend style={cells={align=left},anchor = south west,at={(0,1.2)},font=\normalsize},
   clip mode=individual,
    legend columns=2,
   xmin=0,
   xmax=15,
  ymin=-0.5,
  ymax=1.5,
   ]

\node[align=center] at (rel axis cs:0.1,0.9) {(\textit{c})};

\addplot+[dashed,mark=none, draw =black, draw opacity = 1] 
table {
0 0.5
35 0.5
};

\addplot+[dashed,mark=none, draw =blue, draw opacity = 1] 
table {
0 0.65
35 0.65
};

 \node[align=center, fill=white, fill opacity=0.0, anchor=west,text opacity=1] at (10,1) {$m_w$};

 \node[align=center, fill=white, fill opacity=0.0, anchor=west,text opacity=1] at (1,0) {$m_a$};

\addplot+[xbar,mark=none,bar width=0.035, draw =black, fill = black, fill opacity = 0.8, draw opacity = 0] 
table {
2.02757502027575e-05	-0.186767096615545
0	-0.174155457604546
0.000223033252230332	-0.161543818593546
0.000790754257907543	-0.148932179582547
0.00214922952149229	-0.136320540571548
0.00802919708029197	-0.123708901560548
0.0292376317923764	-0.111097262549549
0.123925385239253	-0.0984856235385497
0.314416058394161	-0.0858739845275503
0.470012165450121	-0.0732623455165510
0.617680454176804	-0.0606507065055517
0.795742092457420	-0.0480390674945524
0.897546634225466	-0.0354274284835531
0.968734793187350	-0.0228157894725537
0.958373884833738	-0.0102041504615544
0.865470397404704	0.00240748854944489
0.710421735604219	0.0150191275604443
0.604927007299268	0.0276307665714435
0.567599351175995	0.0402424055824429
0.499391727493918	0.0528540445934422
0.387226277372261	0.0654656836044415
0.361759935117598	0.0780773226154409
0.325669099756692	0.0906889616264402
0.292842660178427	0.103300600637440
0.249026763990267	0.115912239648439
0.214355231143552	0.128523878659438
0.183008921330091	0.141135517670438
0.171553122465531	0.153747156681437
0.148580697485807	0.166358795692436
0.129440389294405	0.178970434703436
0.117558799675588	0.191582073714435
0.105312246553122	0.204193712725434
0.0927210056772107	0.216805351736433
0.0846512570965123	0.229416990747433
0.0782238442822382	0.242028629758432
0.0735604217356040	0.254640268769431
0.0690186536901871	0.267251907780431
0.0665450121654499	0.279863546791430
0.0643552311435521	0.292475185802429
0.0605636658556371	0.305086824813429
0.0582725060827249	0.317698463824428
0.0527169505271693	0.330310102835427
0.0546228710462291	0.342921741846427
0.0528994322789942	0.355533380857426
0.0499594484995943	0.368145019868425
0.0500202757502026	0.380756658879425
0.0503244120032445	0.393368297890424
0.0496553122465530	0.405979936901423
0.0491686942416873	0.418591575912423
0.0468369829683697	0.431203214923422
0.0479724249797241	0.443814853934421
0.0480535279805351	0.456426492945421
0.0448094079480944	0.469038131956420
0.0415044606650445	0.481649770967419
0.0431873479318733	0.494261409978418
0.0435523114355234	0.506873048989418
0.0430251419302513	0.519484688000417
0.0427007299270072	0.532096327011416
0.0413625304136256	0.544707966022416
0.0415247364152472	0.557319605033415
0.0396593673965935	0.569931244044414
0.0379359286293592	0.582542883055414
0.0401054339010546	0.595154522066413
0.0403081914030818	0.607766161077412
0.0402068126520680	0.620377800088412
0.0395377128953774	0.632989439099411
0.0391930251419301	0.645601078110410
0.0386861313868612	0.658212717121410
0.0388483373884832	0.670824356132409
0.0377737226277375	0.683435995143408
0.0373884833738847	0.696047634154408
0.0365977291159772	0.708659273165407
0.0376926196269265	0.721270912176406
0.0357664233576641	0.733882551187406
0.0378751013787509	0.746494190198405
0.0362327656123279	0.759105829209404
0.0369829683698296	0.771717468220403
0.0382400648824005	0.784329107231403
0.0382603406326033	0.796940746242402
0.0372060016220603	0.809552385253402
0.0370032441200323	0.822164024264401
0.0380981346309812	0.834775663275400
0.0372668288726686	0.847387302286400
0.0377737226277371	0.859998941297399
0.0372668288726682	0.872610580308398
0.0379359286293596	0.885222219319397
0.0376115166261150	0.897833858330397
0.0396999188969991	0.910445497341396
0.0392335766423352	0.923057136352395
0.0392944038929448	0.935668775363395
0.0390510948905108	0.948280414374394
0.0390510948905108	0.960892053385393
0.0447485806974857	0.973503692396393
0.0933090024330897	0.986115331407392
9.47471613949713	0.998726970418391
0.773783454987832	1.01133860942939
0.0463503649635045	1.02395024844039
0.00910381184103809	1.03656188745139
0.00168288726682887	1.04917352646239
0.000587996755879966	1.06178516547339
};

\addplot+[xbar,mark=none,bar width=0.035, draw =black, fill = blue, fill opacity = 0.8, draw opacity = 0] 
table {
3.16856780735108e-05	-0.410889793993891
3.16856780735108e-05	-0.395027355071458
3.16856780735108e-05	-0.379164916149025
0	-0.363302477226592
0.000221799746514575	-0.347440038304159
0.000285171102661597	-0.331577599381727
0.000633713561470216	-0.315715160459294
0.00243979721166033	-0.299852721536861
0.0216096324461344	-0.283990282614428
0.0335551330798476	-0.268127843691995
0.0529467680608365	-0.252265404769562
0.0660963244613435	-0.236402965847129
0.0511089987325729	-0.220540526924697
0.0728770595690748	-0.204678088002264
0.106400506970849	-0.188815649079831
0.170659062103929	-0.172953210157398
0.191666666666667	-0.157090771234965
0.291318124207858	-0.141228332312532
0.394550063371356	-0.125365893390100
0.362230671736375	-0.109503454467667
0.409125475285171	-0.0936410155452338
0.404277566539924	-0.0777785766228009
0.440050697084918	-0.0619161377003681
0.448542458808619	-0.0460536987779353
0.498225602027884	-0.0301912598555024
0.504372623574145	-0.0143288209330696
0.539036755386566	0.00153361798936327
0.509949302915083	0.0173960569117961
0.499588086185045	0.0332584958342290
0.463846641318124	0.0491209347566618
0.504752851711027	0.0649833736790946
0.469391634980989	0.0808458126015275
0.448162230671737	0.0967082515239603
0.424619771863118	0.112570690446393
0.416983523447402	0.128433129368826
0.402693282636249	0.144295568291259
0.385297845373891	0.160158007213692
0.388719898605830	0.176020446136125
0.384474017743980	0.191882885058557
0.357160963244614	0.207745323980990
0.378010139416984	0.223607762903423
0.381178707224335	0.239470201825856
0.372433460076046	0.255332640748289
0.356653992395437	0.271195079670722
0.342363751584284	0.287057518593154
0.333903675538657	0.302919957515587
0.325633713561470	0.318782396438020
0.322750316856781	0.334644835360453
0.302281368821293	0.350507274282886
0.299588086185044	0.366369713205319
0.304816223067174	0.382232152127752
0.301806083650190	0.398094591050184
0.311565272496832	0.413957029972617
0.305291508238271	0.429819468895050
0.283555133079848	0.445681907817483
0.271546261089987	0.461544346739916
0.263276299112801	0.477406785662349
0.276520912547529	0.493269224584782
0.265462610899873	0.509131663507215
0.266951837769328	0.524994102429647
0.253231939163498	0.540856541352080
0.240525982256020	0.556718980274513
0.238656527249683	0.572581419196946
0.237167300380228	0.588443858119379
0.240272496831432	0.604306297041812
0.235868187579214	0.620168735964244
0.248225602027884	0.636031174886677
0.245690747782003	0.651893613809110
0.242363751584284	0.667756052731543
0.252851711026616	0.683618491653976
0.244391634980989	0.699480930576409
0.251108998732573	0.715343369498842
0.263339670468948	0.731205808421274
0.268346007604563	0.747068247343707
0.256812420785801	0.762930686266140
0.271197718631179	0.778793125188573
0.265304182509506	0.794655564111006
0.267965779467681	0.810518003033439
0.285614702154626	0.826380441955872
0.319486692015209	0.842242880878305
0.349081115335868	0.858105319800737
0.369169835234474	0.873967758723170
0.398859315589354	0.889830197645603
0.472116603295311	0.905692636568036
0.575982256020279	0.921555075490469
0.715525982256021	0.937417514412902
0.961121673003803	0.953279953335334
1.66596958174905	0.969142392257767
3.40716096324461	0.985004831180200
4.71045627376426	1.00086727010263
1.88254119138150	1.01672970902507
0.324366286438530	1.03259214794750
0.0580798479087453	1.04845458686993
0.0142268694550063	1.06431702579236
0.00405576679340938	1.08017946471480
0.00114068441064639	1.09604190363723
0.000538656527249683	1.11190434255966
0.000221799746514575	1.12776678148210
0.000158428390367554	1.14362922040453
6.33713561470216e-05	1.15949165932696
};

 \node[align=center, fill=white, fill opacity=0.7, anchor=east] at (rel axis cs:0.99,0.1) {$\overline{c} \approx 0.5$};

\nextgroupplot  [width=4.6cm, height=4.6cm,
    xlabel={PDF},
    grid=both,
    legend style={cells={align=left},anchor = south west,at={(0,1.2)},font=\normalsize},
   clip mode=individual, xmin=0,
   xmax=5,
  ymin=-0.5,
  ymax=1.5,
   ]

 \node[align=center] at (rel axis cs:0.1,0.9) {(\textit{d})};

\addplot+[dashed,mark=none, draw =black, draw opacity = 1] 
table {
0 0.5
35 0.5
};

\addplot+[dashed,mark=none, draw =blue, draw opacity = 1] 
table {
0 0.65
35 0.65
};

 \node[align=center, fill=white, fill opacity=0.0, anchor=west,text opacity=1] at (1.2,1) {$m_w$};

 \node[align=center, fill=white, fill opacity=0.0, anchor=west,text opacity=1] at (3.2,0) {$m_a$};

\addplot+[xbar,mark=none,bar width=0.035, draw =black, fill = black, fill opacity = 0.8, draw opacity = 0] 
table {
1.97005516154452e-05	-0.111684823781449
0.00126083530338849	-0.0999238567188664
0.00819542947202522	-0.0881628896562837
0.0209022852639874	-0.0764019225937011
0.0546099290780142	-0.0646409555311185
0.315484633569740	-0.0528799884685358
1.94952718676123	-0.0411190214059532
2.63695823483058	-0.0293580543433706
2.68264381402679	-0.0175970872807879
3.06767139479905	-0.00583612021820531
2.34074074074074	0.00592484684437733
1.76463750985028	0.0176858139069600
1.33981481481481	0.0294467809695426
1.08400315208826	0.0412077480321252
0.855476753349094	0.0529687150947079
0.655279747832939	0.0647296821572905
0.532742316784870	0.0764906492198731
0.399566587864460	0.0882516162824558
0.287825059101655	0.100012583345038
0.211997635933806	0.111773550407621
0.184515366430260	0.123534517470204
0.147951142631994	0.135295484532786
0.115701339637510	0.147056451595369
0.101122931442080	0.158817418657952
0.0766351457840819	0.170578385720534
0.0626280535855004	0.182339352783117
0.0517927501970055	0.194100319845699
0.0440701339637510	0.205861286908282
0.0367021276595745	0.217622253970865
0.0337273443656422	0.229383221033447
0.0304767533490938	0.241144188096030
0.0306146572104019	0.252905155158613
0.0279156816390859	0.264666122221195
0.0255516154452325	0.276427089283778
0.0227738376674547	0.288188056346361
0.0200157604412924	0.299949023408943
0.0196414499605989	0.311709990471526
0.0183215130023641	0.323470957534109
0.0176910953506698	0.335231924596691
0.0161347517730496	0.346992891659274
0.0169621749408983	0.358753858721856
0.0145981087470448	0.370514825784439
0.0135539795114263	0.382275792847022
0.0144405043341214	0.394036759909604
0.0140464933018125	0.405797726972187
0.0141449960598897	0.417558694034770
0.0137903861308117	0.429319661097352
0.0124704491725768	0.441080628159935
0.0133569739952719	0.452841595222518
0.0131402679275020	0.464602562285100
0.0120370370370370	0.476363529347683
0.0126477541371158	0.488124496410265
0.0125689519306541	0.499885463472848
0.0113278171788810	0.511646430535431
0.0122734436564224	0.523407397598013
0.00992907801418440	0.535168364660596
0.0105988967691095	0.546929331723179
0.0106579984239559	0.558690298785761
0.0119188337273444	0.570451265848344
0.0113869188337273	0.582212232910927
0.0100078802206462	0.593973199973509
0.0103821907013396	0.605734167036092
0.0103033884948779	0.617495134098674
0.0102245862884161	0.629256101161257
0.0111308116627266	0.641017068223840
0.0103624901497242	0.652778035286422
0.00977147360126083	0.664539002349005
0.0113475177304965	0.676299969411588
0.0108156028368794	0.688060936474170
0.0123128447596533	0.699821903536753
0.0136918833727344	0.711582870599335
0.0152285263987392	0.723343837661918
0.0154058313632782	0.735104804724501
0.0135539795114263	0.746865771787083
0.0127659574468085	0.758626738849666
0.0128841607565012	0.770387705912249
0.0135933806146572	0.782148672974831
0.0133569739952719	0.793909640037414
0.0125295508274232	0.805670607099997
0.0126083530338849	0.817431574162579
0.0121749408983452	0.829192541225162
0.0124113475177305	0.840953508287744
0.0120961386918834	0.852714475350327
0.0122931442080378	0.864475442412910
0.0132584712371946	0.876236409475492
0.0124704491725768	0.887997376538075
0.0131599684791174	0.899758343600658
0.0125492513790386	0.911519310663240
0.0120567375886525	0.923280277725823
0.0123916469661151	0.935041244788405
0.0123325453112687	0.946802211850988
0.0132781717888101	0.958563178913571
0.0165681639085894	0.970324145976154
0.0224586288416076	0.982085113038736
0.794661150512214	0.993846080101319
0.854491725768321	1.00560704716390
0.00839243498817967	1.01736801422648
0.00159574468085106	1.02912898128907
0.000413711583924350	1.04088994835165
0.000137903861308117	1.05265091541423
};

\addplot+[xbar,mark=none,bar width=0.035, draw =black, fill = blue, fill opacity = 0.8, draw opacity = 0] 
table {
8.56164383561644e-05	-0.261989176851835
5.70776255707763e-05	-0.247808394293450
2.85388127853881e-05	-0.233627611735065
0.000171232876712329	-0.219446829176680
0.000371004566210044	-0.205266046618295
0.00125570776255708	-0.191085264059910
0.00285388127853880	-0.176904481501525
0.00776255707762560	-0.162723698943140
0.0404109589041094	-0.148542916384755
0.298715753424659	-0.134362133826370
0.424343607305934	-0.120181351267985
0.510530821917810	-0.106000568709600
0.531107305936071	-0.0918197861512155
0.841638127853885	-0.0776390035928305
1.18319063926940	-0.0634582210344456
1.41646689497717	-0.0492774384760606
1.73153538812785	-0.0350966559176757
1.83473173515982	-0.0209158733592907
2.14457762557077	-0.00673509080090578
1.98681506849316	0.00744569175747923
1.73858447488585	0.0216264743158641
1.66766552511415	0.0358072568742490
1.49788812785388	0.0499880394326340
1.30131278538813	0.0641688219910191
1.25074200913242	0.0783496045494040
1.17317351598173	0.0925303871077889
1.07682648401826	0.106711169666174
0.918721461187218	0.120891952224559
0.762756849315071	0.135072734782944
0.628053652968034	0.149253517341329
0.587414383561642	0.163434299899714
0.576312785388130	0.177615082458099
0.475998858447490	0.191795865016484
0.409988584474884	0.205976647574869
0.339897260273971	0.220157430133254
0.281192922374430	0.234338212691639
0.276369863013700	0.248518995250024
0.251312785388129	0.262699777808408
0.219549086757988	0.276880560366793
0.197231735159818	0.291061342925178
0.179880136986302	0.305242125483563
0.197545662100457	0.319422908041948
0.173915525114153	0.333603690600333
0.167665525114156	0.347784473158718
0.155450913242010	0.361965255717103
0.137328767123288	0.376146038275488
0.132905251141551	0.390326820833873
0.131449771689498	0.404507603392258
0.117808219178083	0.418688385950643
0.101398401826484	0.432869168509028
0.0969463470319624	0.447049951067413
0.0905821917808223	0.461230733625798
0.0920091324200917	0.475411516184183
0.0872431506849318	0.489592298742568
0.0802796803652959	0.503773081300953
0.0791381278538816	0.517953863859338
0.0746860730593610	0.532134646417723
0.0706621004566213	0.546315428976108
0.0680365296803645	0.560496211534493
0.0633847031963473	0.574676994092878
0.0668664383561646	0.588857776651263
0.0562785388127856	0.603038559209647
0.0559646118721455	0.617219341768032
0.0579908675799089	0.631400124326417
0.0605308219178084	0.645580906884802
0.0580194063926943	0.659761689443187
0.0564497716894979	0.673942472001572
0.0519691780821912	0.688123254559957
0.0469178082191783	0.702304037118342
0.0472888127853883	0.716484819676727
0.0468321917808221	0.730665602235112
0.0419805936073061	0.744846384793497
0.0421232876712324	0.759027167351882
0.0437785388127856	0.773207949910267
0.0466609589041098	0.787388732468652
0.0462043378995436	0.801569515027037
0.0446632420091319	0.815750297585422
0.0410958904109591	0.829931080143807
0.0479737442922376	0.844111862702192
0.0520833333333335	0.858292645260577
0.0571061643835610	0.872473427818961
0.0609018264840185	0.886654210377347
0.0693207762557080	0.900834992935731
0.0801940639269409	0.915015775494116
0.0914383561643825	0.929196558052501
0.122602739726028	0.943377340610886
0.211586757990868	0.957558123169271
0.281250000000001	0.971738905727656
0.332448630136983	0.985919688286041
0.394977168949773	1.00010047084443
0.304394977168951	1.01428125340281
0.154680365296804	1.02846203596120
0.0920947488584464	1.04264281851958
0.0402968036529682	1.05682360107797
0.0327910958904111	1.07100438363635
0.0170091324200914	1.08518516619474
0.00308219178082193	1.09936594875312
0.00208333333333331	1.11354673131151
0.000428082191780824	1.12772751386989
0.000256849315068494	1.14190829642828
};

 \node[align=center, fill=white, fill opacity=0.7, anchor=east] at (rel axis cs:0.99,0.1) {$ \overline{c}  \approx 0.9$};

\end{groupplot}

\end{tikzpicture}

%% file: Figures/fig7.tex
\pgfplotstableread{
interval WRL1 WRL2 WRL3 CBR1 CBR2 CBR3
0.125000000000000	0.438764643237487	0.300925925925926	0.362652232746955	0.668041237113402	0.621915103652517	1.92171294747407
0.375000000000000	0.690095846645367	0.587962962962963	0.197564276048715	0.614432989690722	0.465942744323791	0.176647708263633
0.625000000000000	0.685835995740149	0.566358024691358	0.132611637347767	0.507216494845361	0.444225074037512	0.121779859484778
0.875000000000000	0.464323748668797	0.450617283950617	0.167794316644114	0.437113402061856	0.319842053307009	0.0802944128471061
1.12500000000000	0.430244941427050	0.324074074074074	0.127198917456022	0.317525773195876	0.282329713721619	0.0521913683506189
1.37500000000000	0.349307774227902	0.259259259259259	0.132611637347767	0.280412371134021	0.250740375123396	0.0776179324188692
1.62500000000000	0.166134185303514	0.216049382716049	0.108254397834912	0.177319587628866	0.199407699901283	0.0695884911341586
1.87500000000000	0.195953141640043	0.166666666666667	0.0974289580514208	0.148453608247423	0.163869693978282	0.0588825694212111
2.12500000000000	0.115015974440895	0.125000000000000	0.0947225981055480	0.156701030927835	0.108588351431392	0.0602208096353295
2.37500000000000	0.0809371671991480	0.106481481481481	0.108254397834912	0.119587628865979	0.106614017769003	0.0508531281365005
2.62500000000000	0.0511182108626198	0.0848765432098766	0.0730717185385656	0.0824742268041237	0.0927936821322804	0.0495148879223821
2.87500000000000	0.0340788072417465	0.0725308641975309	0.0920162381596752	0.0453608247422680	0.0789733464955578	0.0468384074941452
3.12500000000000	0.0383386581469649	0.0540123456790123	0.108254397834912	0.0701030927835052	0.102665350444225	0.0495148879223821
3.37500000000000	0.0298189563365282	0.0509259259259259	0.0838971583220568	0.0494845360824742	0.0315893385982231	0.0428236868517899
3.62500000000000	0.0170394036208733	0.0555555555555556	0.0568335588633288	0.0371134020618557	0.0513326752221125	0.0334560053529609
3.87500000000000	0.0425985090521832	0.0370370370370370	0.0595399188092016	0.0206185567010309	0.0454096742349457	0.0347942455670793
4.12500000000000	0.0212992545260916	0.0401234567901235	0.0541271989174560	0.00824742268041237	0.0473840078973347	0.0294412847106056
4.37500000000000	0.0170394036208733	0.0370370370370370	0.0811907983761840	0.0164948453608247	0.0414610069101678	0.0294412847106056
4.62500000000000	0.00851970181043664	0.0385802469135802	0.0487144790257104	0.0288659793814433	0.0296150049358342	0.0254265640682503
4.87500000000000	0.0212992545260916	0.0401234567901235	0.0487144790257104	0.0164948453608247	0.0315893385982231	0.0267648042823687
5.12500000000000	0.0212992545260916	0.0262345679012346	0.0460081190798376	0.0288659793814433	0.0177690029615005	0.0281030444964871
5.37500000000000	0.0170394036208733	0.0200617283950617	0.0405953991880920	0.0164948453608247	0.0138203356367226	0.0133824021411843
5.62500000000000	0	0.0185185185185185	0.0297699594046008	0.0123711340206186	0.0375123395853899	0.0120441619270659
5.87500000000000	0.00425985090521832	0.0200617283950617	0.0270635994587280	0.0123711340206186	0.0256663376110563	0.0214118434258949
6.12500000000000	0.00425985090521832	0.0246913580246914	0.0162381596752368	0.0123711340206186	0.0276406712734452	0.0173971227835396
6.37500000000000	0.00425985090521832	0.0169753086419753	0.0378890392422192	0.00824742268041237	0.0197433366238894	0.0267648042823687
6.62500000000000	0.00851970181043664	0.0185185185185185	0.0351826792963464	0.0164948453608247	0.0177690029615005	0.0267648042823687
6.87500000000000	0.00425985090521832	0.0216049382716049	0.0189445196211096	0.0123711340206186	0.0217176702862784	0.0107059217129475
7.12500000000000	0.00425985090521832	0.00771604938271605	0.0270635994587280	0	0.0157946692991116	0.0133824021411843
7.37500000000000	0.00851970181043664	0.00925925925925926	0.0243572395128552	0.00412371134020619	0.0177690029615005	0.0187353629976581
7.62500000000000	0	0.00771604938271605	0.0108254397834912	0.00412371134020619	0.0138203356367226	0.00802944128471061
7.87500000000000	0	0.0108024691358025	0.0243572395128552	0.00412371134020619	0.00789733464955577	0.00802944128471061
8.12500000000000	0.00425985090521832	0.0154320987654321	0.0135317997293640	0	0.00789733464955577	0.0147206423553028
8.37500000000000	0	0.00617283950617284	0.0189445196211096	0.0123711340206186	0.00987166831194472	0.0147206423553028
8.62500000000000	0.00851970181043664	0.00617283950617284	0.0189445196211096	0	0.00592300098716683	0.0107059217129475
8.87500000000000	0	0.00771604938271605	0.0297699594046008	0	0.0118460019743337	0.00669120107059217
9.12500000000000	0	0.00771604938271605	0.0135317997293640	0.00824742268041237	0.0236920039486673	0.0120441619270659
9.37500000000000	0	0.00617283950617284	0.0189445196211096	0	0	0.0147206423553028
9.62500000000000	0	0.00462962962962963	0.0243572395128552	0.00412371134020619	0.0118460019743337	0.0107059217129475
9.87500000000000	0	0.00617283950617284	0.0216508795669824	0	0.00197433366238894	0.00936768149882904
10.1250000000000	0.0127795527156550	0.123456790123457	1.25845737483085	0.0412371134020619	0.173741362290227	0.659752425560389
}\mydata

\begin{tikzpicture}

\begin{groupplot}[
     group style = {group size = 1 by 3,horizontal sep=0.3cm,vertical sep=1.4cm},
    width = 1\textwidth]

\nextgroupplot[height = 5cm,
     width = 7cm,
     ylabel={PDF},
     xlabel={$t_\text{ch,a}$ (ms)},
    xtick={0,1,2,3,4,5,6,7,8,9,10},
     grid=both,
     xmin=-0.5,
     xmax=11,
     ymin=0,
     ymax=1,
    legend columns=1,
   legend style={cells={align=left},anchor = south west,at={(0,1.15)},font=\normalsize},
   clip mode=individual,
     ]

\addplot+[xbar,fill=black,mark=none,bar width=0.05,color=black,draw=none,bar shift=-0.0625,draw=none] 
table[row sep=crcr]{%
0 -1	\\
};
\addlegendentry{State-of-the-art PD \,};

\addplot+[xbar,fill=blue,color=blue,mark=none,bar width=0.05, draw=none,bar shift=0.0625,draw=none] 
table[row sep=crcr]{%
0 -1	\\
};
\addlegendentry{NexGen PD};

\addplot[ybar, fill=black,bar width=0.125,draw=none,bar shift=-0.0625,draw=none] table[x=interval,y=WRL1]{\mydata}; 

\addplot[ybar, fill=blue,bar width=0.125,bar shift=0.0625,draw=none] table[x=interval,y=CBR1]{\mydata};

 \node[align=center,anchor=west,fill=white,fill opacity =0.6,text=black,text opacity=1] at (rel axis cs:0.1,0.9) {(\textit{a}) \textcolor{gray}{$ \, \, \, \overline{c} \approx 0.02$}};

\node[align=center,fill=white,fill opacity =0.6,text=gray,draw=none,anchor=east] at (8.9,0.2) {\small{$ t_\text{ch,a} > 10$ ms}};

 \draw[gray,line width=1.3,arrows={->[scale=1,gray,line width=1.5]}]   (9,0.2)   -- (10,0.09);

\nextgroupplot[height = 5cm,
     width = 7cm,
     ylabel={PDF},
    xlabel={$t_\text{ch,a}$ (ms)},
    xtick={0,1,2,3,4,5,6,7,8,9,10},
     grid=both,
     xmin=-0.5,
     xmax=11,
     ymin=0,
     ymax=0.8,
    legend columns=4,
   legend style={cells={align=left},anchor = south west,at={(0,1.2)},font=\normalsize},
   clip mode=individual,
     ]
     
 \node[align=center,anchor=west,fill=white,fill opacity =0.6,text=black,text opacity=1] at (rel axis cs:0.1,0.9) {(\textit{b}) \textcolor{gray}{$ \, \, \, \overline{c} \approx 0.1$}};
 
\addplot[ybar, fill=black,bar width=0.125,draw=none,bar shift=-0.0625] table[x=interval,y=WRL2]{\mydata};

\addplot[ybar, fill=blue,bar width=0.125,bar shift=0.0625,draw=none] table[x=interval,y=CBR2]{\mydata};

\node[align=center,fill=white,fill opacity =0.6,text=gray,draw=none,anchor=east] at (8.4,0.25) {\small{$ t_\text{ch,a} > 10$ ms}};

 \draw[gray,line width=1.3,arrows={->[scale=1,gray,line width=1.5]}]   (8.5,0.25)   -- (9.7,0.2);

\nextgroupplot[height = 5cm,
     width = 7cm,
     ylabel={PDF},
   xlabel={$t_\text{ch,a}$ (ms)},
    xtick={0,1,2,3,4,5,6,7,8,9,10},
     grid=both,
     xmin=-0.5,
     xmax=11,
     ymin=0,
     ymax=0.6,
    legend columns=4,
   legend style={cells={align=left},anchor = south west,at={(0,1.2)},font=\normalsize},
   clip mode=individual,
     ]

 \node[align=center,anchor=west,fill=white,fill opacity =0.6,text=black,text opacity=1] at (rel axis cs:0.1,0.9) {(\textit{c}) \textcolor{gray}{$ \, \, \, \overline{c} \approx 0.5$}};

\addplot[ybar, fill=black,bar width=0.125,draw=none,bar shift=-0.0625] table[x=interval,y=WRL3]{\mydata};

\addplot[ybar, fill=blue,bar width=0.125,bar shift=0.0625,draw=none] table[x=interval,y=CBR3]{\mydata};

\node[align=center,fill=white,fill opacity =0.6,text=gray,draw=none,anchor=east] at (8.4,0.25) {\small{$ t_\text{ch,a} > 10$ ms}};

 \draw[gray,line width=1.3,arrows={->[scale=1,gray,line width=1.5]}]   (8.5,0.25)   -- (9.7,0.2);
 
\end{groupplot}

\end{tikzpicture}

%% file: Figures/fig8.tex
\begin{tikzpicture}

\begin{groupplot}[
     group style = {group size = 3 by 2,horizontal sep=1.4cm,vertical sep=1.4cm},
    width = 1\textwidth]

\nextgroupplot  [width=4.6cm, height=4.6cm,
    xlabel={$\langle \overline{c} \rangle_\text{NexGen}$},
    ylabel={$\langle \overline{c} \rangle_\text{SOTA}$},
    grid=both,
    legend style={cells={align=left},anchor = south west,at={(0,1.2)},font=\normalsize},
   clip mode=individual,
    legend columns=2,
   xmin=0,
   xmax=0.6,
  ymin=0,
  ymax=0.6,
   ]

\node[align=center, fill=white, fill opacity=0.7, anchor=east] at (rel axis cs:0.99,0.1) {$\epsilon = 0.8$ to 6.4\%};

\addplot+[densely dashed,mark=none, draw =black, draw opacity = 1] 
table {
0 0
1 1
};

\addplot+[dashed,mark=none, draw =gray, draw opacity = 0.5] 
table {
0 0
1 1.05
};

\addplot+[dashed,mark=none, draw =gray, draw opacity = 0.5] 
table {
0 0
1 0.95
};

\addplot [fill=black,mark=square*,only marks,fill opacity=1,draw opacity=1, mark size=3,clip mode=individual]
table[row sep=crcr]{%
0.337126205781378	0.347800030605586\\
0.259919933520652	0.265775136752869\\
0.208459150208594	0.206052169029167\\
0.149010171567600	0.138109381572801\\
0.121844975342684	0.114743128538792\\
};

\node[align=center] at (rel axis cs:0.1,0.9) {(\textit{a})};

\nextgroupplot  [width=4.6cm, height=4.6cm,
       xlabel={$(y_{90}/d_1)_\text{NexGen}$},
    ylabel={$(y_{90}/d_1)_\text{SOTA}$},
    grid=both,
    legend style={cells={align=left},anchor = south west,at={(0,1.2)},font=\normalsize},
   clip mode=individual,
    legend columns=2,
   xmin=0,
   xmax=6,
  ymin=0,
  ymax=6,
   ]

 \node[align=center, fill=white, fill opacity=0.7, anchor=east] at (rel axis cs:0.99,0.1) {$\epsilon = 0.1$ to 0.85\%};

\addplot+[densely dashed,mark=none, draw =black, draw opacity = 1] 
table {
0 0
6 6
};

\addplot+[dashed,mark=none, draw =gray, draw opacity = 0.5] 
table {
0 0
6 5.7
};

\addplot+[dashed,mark=none, draw =gray, draw opacity = 0.5] 
table {
0 0
6 6.3
};

\addplot [fill=black,mark=square*,only marks,fill opacity=1,draw opacity=1, mark size=3,clip mode=individual]
table[row sep=crcr]{%
3.04946086815363	3.03012732821286\\
3.68014972975665	3.66036568107577\\
4.23174471635950	4.22864963685373\\
4.54895756548524	4.56785874666958\\
4.85686492238828	4.89951838380286\\
};

\node[align=center] at (rel axis cs:0.1,0.9) {(\textit{b})};

\nextgroupplot  [width=4.6cm, height=4.6cm,
  xlabel={$(d_\text{eq}/d_1)_\text{NexGen}$},
    ylabel={$(d_\text{eq}/d_1)_\text{SOTA}$},
    grid=both,
    legend style={cells={align=left},anchor = south west,at={(0,1.2)},font=\normalsize},
   clip mode=individual,
    legend columns=2,
   xmin=0,
   xmax=6,
  ymin=0,
  ymax=6,
   ]

 \node[align=center, fill=white, fill opacity=0.7, anchor=east] at (rel axis cs:0.99,0.1) {$\epsilon = 0.1$ to 2.7\%};   

\addplot+[densely dashed,mark=none, draw =black, draw opacity = 1] 
table {
0 0
6 6
};

\addplot+[dashed,mark=none, draw =gray, draw opacity = 0.5] 
table {
0 0
6 6.3
};

\addplot+[dashed,mark=none, draw =gray, draw opacity = 0.5] 
table {
0 0
6 5.7
};

\addplot [fill=black,mark=square*,only marks,fill opacity=1,draw opacity=1, mark size=3,clip mode=individual]
table[row sep=crcr]{%
2.02140769599421	1.97624895072161\\
2.72360545665226	2.68753149162235\\
3.34959880888749	3.35732720711561\\
3.87111661819855	3.93699460005513\\
4.26508033567713	4.33733231611199\\
};

\node[align=center] at (rel axis cs:0.1,0.9) {(\textit{c})};

\nextgroupplot  [width=4.6cm, height=4.6cm,
     xlabel={$(F_\text{TSL,max} d_1/U_1)_\text{NexGen}$},
       ylabel={$(F_\text{TSL,max} d_1/U_1)_\text{SOTA}$},
    grid=both,
    legend style={cells={align=left},anchor = south west,at={(0,1.2)},font=\normalsize},
   clip mode=individual,
    xmin=0,
   xmax=2,
  ymin=0,
  ymax=2,
   ]

 \node[align=center] at (rel axis cs:0.1,0.9) {(\textit{d})};

\addplot+[densely dashed,mark=none, draw =black, draw opacity = 1] 
table {
0 0
2 2
};

\addplot+[dashed,mark=none, draw =gray, draw opacity = 0.5] 
table {
0 0
2 2.1
};

\addplot+[dashed,mark=none, draw =gray, draw opacity = 0.5] 
table {
0 0
2 1.9
};

\addplot [fill=black,mark=square*,only marks,fill opacity=1,draw opacity=1, mark size=3,clip mode=individual]
table[row sep=crcr]{%
1.014853714	1.157476852	\\
0.799544696	0.993475514	\\
0.772669244	0.702121183	\\
0.533844205	0.451690835	\\
0.408934434	0.277916606	\\
};

\nextgroupplot  [width=4.6cm, height=4.6cm,
     xlabel={$(F_\text{TWL,max} d_1/U_1)_\text{NexGen}$},
       ylabel={$(F_\text{TWL,max} d_1/U_1)_\text{SOTA}$},
    grid=both,
    legend style={cells={align=left},anchor = north west,at={(1.2,1)},font=\normalsize},
   clip mode=individual,
    xmin=0,
   xmax=2,
  ymin=0,
  ymax=2,
   ]

 \node[align=center] at (rel axis cs:0.1,0.9) {(\textit{e})};

\addplot+[densely dashed,mark=none, draw =black, draw opacity = 1] 
table {
0 0
2 2
}; \addlegendentry{1:1}

\addplot+[dashed,mark=none, draw =gray, draw opacity = 0.5] 
table {
0 0
2 2.1
}; \addlegendentry{$\pm 5$ \%}

\addplot+[dashed,mark=none, draw =gray, draw opacity = 0.5] 
table {
0 0
2 1.9
};

\addplot [fill=black,mark=square*,only marks,fill opacity=1,draw opacity=1, mark size=3,clip mode=individual]
table[row sep=crcr]{%
1.194736001	0.470015007	\\
0.969043285	0.472763632	\\
0.779388107	0.368621256	\\
0.480398704	0.293186749	\\
0.386334622	0.27944362	\\
};

\end{groupplot}

\end{tikzpicture}

%% file: Figures/fig9.tex
\pgfplotstableread{
t S1 S2
0.0694444430555556	1.08106496909991	1.17888681493332
0.138888886111111	1.10711214546520	1.12140735164761
0.208333329166667	1.11823469889113	1.08626807807269
0.277777772222222	1.13229666740308	1.05087118439978
0.347222215277778	1.13093683725327	1.02748865730507
0.416666658333333	1.13524387521201	1.01355507489226
0.486111101388889	1.14735532517669	1.00304870210941
0.555555544444444	1.13942778814859	0.996930494259249
0.624999987500000	1.13821056482825	0.994075653968350
0.694444430555556	1.14448526532468	0.995631614685704
0.763888873611111	1.14773171369380	0.994401182293005
0.833333316666667	1.13855812027055	0.990705034527291
0.902777759722222	1.13757312009326	0.990564367486339
0.972222202777778	1.13404457515438	0.993273690148667
1.04166664583333	1.12587624298801	0.987198598632900
1.11111108888889	1.10343086315706	0.983017931059461
1.18055553194444	1.09062430230778	0.981705038677238
1.24999997500000	1.05139573512067	0.975454787053006
1.31944441805556	1.01917594292921	0.966591685564646
1.38888886111111	0.990153504952470	0.967458862840096
1.45833330416667	0.982589108812517	0.968534076428985
1.52777774722222	0.972433632301148	0.971910085411842
1.59722219027778	0.988446119360357	0.976901879026021
1.66666663333333	0.993126428747043	0.981744382332370
1.73611107638889	0.995918561369836	0.983155903329513
1.80555551944444	0.991883489441055	0.981530417522952
1.87499996250000	0.989327476322786	0.977673661411630
1.94444440555556	0.976029973929446	0.972960507108455
2.01388884861111	0.962010865393570	0.975549104034487
2.08333329166667	0.946303853072885	0.975773308973323
2.15277773472222	0.938779199819832	0.982678928880307
2.22222217777778	0.929778604845437	0.985602755306233
2.29166662083333	0.925480138881908	0.989870733456815
2.36111106388889	0.928116417046270	0.992800488378720
2.43055550694444	0.931453261001279	0.993994810841365
2.49999995000000	0.928703988354555	0.991496758217556
2.56944439305556	0.930694249788720	0.992059965335545
2.63888883611111	0.935142336032343	0.992947622869831
2.70833327916667	0.937268190845343	0.985929361539249
2.77777772222222	0.940970513505373	0.980994697068138
2.84722216527778	0.947398730643363	0.974376339739038
2.91666660833333	0.946516594408643	0.974839840333747
2.98611105138889	0.940598800622015	0.972043746048455
3.05555549444444	0.940312807690794	0.976920203468138
3.12499993750000	0.941982788212883	0.983223811556180
3.19444438055556	0.943948112933699	0.995441902814381
3.26388882361111	0.946243849106389	1.00176706906962
3.33333326666667	0.946825186236337	1.01014241702517
3.40277770972222	0.942864924447603	1.01379976008994
3.47222215277778	0.942486198113615	1.02189916350539
3.54166659583333	0.943622377115578	1.02492485227131
3.61111103888889	0.938951418996398	1.02785245137650
3.68055548194444	0.944278524385573	1.02669693361480
3.74999992500000	0.951554589777428	1.02781849726316
3.81944436805556	0.958658435992717	1.02598066350973
3.88888881111111	0.967104189095097	1.00685210175692
3.95833325416667	0.983135378689319	0.992860851246868
4.02777769722222	0.996171045592494	0.981994996026021
4.09722214027778	1.00433080576365	0.969255736075863
4.16666658333333	1.00745257056603	0.954326705292794
4.23611102638889	1.00886617050400	0.966288254361365
4.30555546944444	1.00080381936936	0.974434007836286
4.37499991250000	0.991888944347100	0.982122189212476
4.44444435555556	0.990646784313405	0.988816000126762
4.51388879861111	0.988837314051020	0.999942709170677
4.58333324166667	0.983385525095101	0.998811983301259
4.65277768472222	0.986585217126691	1.00161346712835
4.72222212777778	0.987473587539748	1.00480030319406
4.79166657083333	0.980058032407597	1.01040434875660
4.86111101388889	0.977163815114531	1.00938464744824
4.93055545694444	0.974965487978361	1.01393773235999
4.99999990000000	0.970806511755127	1.01400240686158
5.06944434305556	0.965996063895652	1.01489653184602
5.13888878611111	0.964181138727222	1.01339716131756
5.20833322916667	0.955355880018562	1.01645464838010
5.27777767222222	0.945759921012960	1.01847680446306
5.34722211527778	0.937600160841801	1.01967543855914
5.41666655833333	0.938247736116582	1.02097755185777
5.48611100138889	0.940589449354509	1.01981664455428
5.55555544444445	0.952333862069583	1.01814804241332
5.62499988750000	0.963899042157462	1.01499246569004
5.69444433055556	0.973797358812420	1.01077137655311
5.76388877361111	0.980065045858226	1.00649100245639
5.83333321666667	0.989092136090554	1.00674862255438
5.90277765972222	0.990591455980662	1.00816014355152
5.97222210277778	0.989724905191785	1.00454645577533
6.04166654583333	0.994455088005168	1.00559633851777
6.11111098888889	1.00216520806375	1.00565670138592
6.18055543194445	1.00407910081329	1.00749992468115
6.24999987500000	1.00886071559795	1.00218691437576
6.31944431805556	1.01209001997664	1.00055442216486
6.38888876111111	1.01176428415852	0.996246561404963
6.45833320416667	1.00985117068128	0.993368007130148
6.52777764722222	1.01464681236721	0.988897382207926
6.59722209027778	1.01841537317207	0.989719287332265
6.66666653333333	1.02355857030030	0.990931395282847
6.73611097638889	1.02704815162457	0.995888156875333
6.80555541944444	1.02470877620352	0.998911150870360
6.87499986250000	1.01442290146193	1.00075706893650
6.94444430555556	0.999533605533146	0.999614486075122
7.04999999861111	0.979196607179681	1.08058551898368
7.09999999722222	0.988157365024747	1.09186539447305
7.14999999583333	0.978928384416300	1.09686295632975
7.19999999444444	0.982505629060553	1.10233183213288
7.24999999305556	0.980835589433816	1.10084078476172
7.29999999166667	0.979258016075508	1.09586926517075
7.34999999027778	0.985580308948155	1.09256751438242
7.39999998888889	0.985984760641359	1.08197188665926
7.44999998750000	0.975786648837853	1.07040350371622
7.49999998611111	0.974513296561169	1.06542432463533
7.54999998472222	0.973135478576642	1.05911903253289
7.59999998333333	0.968982966689873	1.05399279124553
7.64999998194444	0.970391136197625	1.05206157852028
7.69999998055556	0.968099243269470	1.05355926411604
7.74999997916667	0.961855299677705	1.04954109570382
7.79999997777778	0.967348219532003	1.04743014028177
7.84999997638889	0.973478521374019	1.04447970476445
7.89999997500000	0.974232368508996	1.04223189978916
7.94999997361111	0.982532451336891	1.03748046287523
7.99999997222222	0.984626706439397	1.03307880933433
8.04999997083333	0.980977465158657	1.02946863641854
8.09999996944445	0.975408313571609	1.02629709695885
8.14999996805556	0.972663971192593	1.02347942593349
8.19999996666667	0.973132655179133	1.02164216961792
8.24999996527778	0.972926547160956	1.02173204096632
8.29999996388889	0.970529482675580	1.02089051834041
8.34999996250000	0.967370100862700	1.01647099265636
8.39999996111111	0.963768151490240	1.01379119244956
8.44999995972222	0.958156648940554	1.01227206028202
8.49999995833333	0.952496442783832	1.01266524743127
8.54999995694445	0.947990300359027	1.01512036926707
8.59999995555556	0.949133776350284	1.01929223811036
8.64999995416667	0.951201209176452	1.02424486429286
8.69999995277778	0.949790922120568	1.02688789894804
8.74999995138889	0.952398329720385	1.02659837022905
8.79999995000000	0.957666789472694	1.02611020540480
8.84999994861111	0.958221587083267	1.02443584090821
8.89999994722222	0.959634697536660	1.02199144235829
8.94999994583333	0.965535598331048	1.02150634133000
8.99999994444444	0.968150064424637	1.01864322399777
9.04999994305556	0.969493295789676	1.01464497025935
9.09999994166667	0.972916665269673	1.01345774932169
9.14999994027778	0.971981414844725	1.00814614837826
9.19999993888889	0.973136184426019	1.00206604527948
9.24999993750000	0.973266060711446	0.999418925563009
9.29999993611111	0.975200793854678	0.995584074276219
9.34999993472222	0.977179289659303	0.988506705589812
9.39999993333333	0.977815259948268	0.986406473453651
9.44999993194444	0.978039720050255	0.983019957643532
9.49999993055556	0.981530851070475	0.980859981485992
9.54999992916667	0.979188842836531	0.977010832484684
9.59999992777778	0.980207383488002	0.976018673223667
9.64999992638889	0.984837755403216	0.970289885395904
9.69999992500000	0.984227901541212	0.963872764740631
9.74999992361111	0.983608165787926	0.954855502573729
9.79999992222222	0.990557252907632	0.949965684208575
9.84999992083333	0.992983963066855	0.950338956684027
9.89999991944445	0.996397450655570	0.954104361928873
9.94999991805556	1.00440178259437	0.958530525837515
9.99999991666667	1.01259457631691	0.964891476900038
10.0499999152778	1.01220706500876	0.969381469891350
10.0999999138889	1.00710306816137	0.968763604371108
10.1499999125000	1.00390133538585	0.967958847296809
10.1999999111111	1.00310090219197	0.967425746798353
10.2499999097222	1.00081889115510	0.971729869500930
10.2999999083333	1.00212824175002	0.976600794457614
10.3499999069444	1.01299620461261	0.979450124707985
10.3999999055556	1.02288515438885	0.980203818516148
10.4499999041667	1.02786139249895	0.985348953211981
10.4999999027778	1.03289974535425	0.983465229324235
10.5499999013889	1.04113700758756	0.983385570629063
10.5999999000000	1.03752658802258	0.985714566197512
10.6499998986111	1.03325337589229	0.986371750432679
10.6999998972222	1.03598783638002	0.984667258609069
10.7499998958333	1.04102336583781	0.985678311278556
10.7999998944444	1.03983542133579	0.986038817937474
10.8499998930556	1.04642452527306	0.980792067342031
10.8999998916667	1.04635464618471	0.978890471311138
10.9499998902778	1.04493306553879	0.974622603527517
10.9999998888889	1.04153228323886	0.968585904204956
11.0499998875000	1.04200661402042	0.961946658341993
11.0999998861111	1.04422015766769	0.961122497226563
11.1499998847222	1.04830984895988	0.955332944112098
11.1999998833333	1.04945967759553	0.953374157223133
11.2499998819444	1.04852160377308	0.951656388950328
11.2999998805556	1.04866700874480	0.951037502164764
11.3499998791667	1.04316420699922	0.949373860954063
11.3999998777778	1.04129370614933	0.954043596642172
11.4499998763889	1.04238071419040	0.954698227714031
11.4999998750000	1.04149204982435	0.954578739671274
11.5499998736111	1.04283881043628	0.956113701451313
11.5999998722222	1.04660239931614	0.957382112982123
11.6499998708333	1.05343572713797	0.954470996179727
11.6999998694444	1.05088549333772	0.952411104023814
11.7499998680556	1.05656616912638	0.951063033797831
11.7999998666667	1.05868654065585	0.949777260756534
11.8499998652778	1.06196732856163	0.946165555942757
11.8999998638889	1.06272964588913	0.947145460019901
11.9499998625000	1.06881830261789	0.949816749682346
11.9999998611111	1.07105584514399	0.954392869382540
}\mydata

\begin{tikzpicture}

\begin{groupplot}[
     group style = {group size = 3 by 2,horizontal sep=1.6cm,vertical sep=1.6cm},
    width = 1\textwidth]

\nextgroupplot  [width=5cm, height=5cm,
    xlabel={$t$ (h)},
    ylabel={$\frac{m_w - m_a}{(\overline{m_w - m_a})}$},
    grid=both,
    legend style={cells={align=left},anchor = north west,at={(1.1,1)},font=\normalsize},
   clip mode=individual,
    legend columns=1,
   xmin=0,
   xmax=12,
  ymin=0,
  ymax=1.5,
   ]

\addplot [solid,draw=blue,no marks,fill opacity=1,draw opacity=1, mark size=2,clip mode=individual, line width=2]
table [y=S1,x=t] {\mydata}; \addlegendentry{$S_1$}

\addplot [solid,draw=black,no marks,fill opacity=1,draw opacity=1, mark size=2,clip mode=individual, line width=2]
table [y=S2,x=t] {\mydata}; \addlegendentry{$S_2$}

\node[align=center, color = gray, draw opacity=0.7, anchor=east,fill = white, fill opacity=0.7] at (rel axis cs:0.99,0.1) {$\overline{c} \approx 0.5$};
 
\end{groupplot}

\end{tikzpicture}